\documentclass[10pt]{article}
\usepackage{graphicx}
\usepackage{amsmath}
\usepackage{amssymb}
\usepackage{caption2}
\begin{document}
\bibliographystyle{prsty}
\begin{center}
{\large {\bf \sc{  Analysis of  the $Z_c(4020)$, $Z_c(4025)$, $Y(4360)$ and $Y(4660)$ as   vector tetraquark states
with  QCD sum rules }}} \\[2mm]
Zhi-Gang  Wang \footnote{E-mail: zgwang@aliyun.com.  }     \\
 Department of Physics, North China Electric Power University, Baoding 071003, P. R. China
\end{center}

\begin{abstract}
In this article, we  distinguish the charge conjugations of the interpolating currents,  calculate the contributions of the vacuum condensates up to dimension-10 in the operator product expansion, and study the masses and pole residues of the $J^{PC}=1^{-\pm}$ hidden charmed tetraquark states with the QCD sum rules.
 We suggest a  formula $\mu=\sqrt{M^2_{X/Y/Z}-(2{\mathbb{M}}_c)^2}$ with the effective mass ${\mathbb{M}}_c=1.8\,\rm{GeV}$ to estimate the energy scales of the QCD spectral densities of the hidden charmed tetraquark states, which works very well. The numerical results disfavor assigning  the $Z_c(4020)$, $Z_c(4025)$, $Y(4360)$ as the diquark-antidiquark (with the Dirac spinor structure $C-C\gamma_\mu$) type vector tetraquark states, and favor assigning  the $Z_c(4020)$, $Z_c(4025)$ as the
diquark-antidiquark  type $1^{+-}$ tetraquark states. While the masses of the  tetraquark states with symbolic quark structures $c\bar{c}s\bar{s}$ and $c\bar{c}(u\bar{u}+d\bar{d})/\sqrt{2}$ favor  assigning the $Y(4660)$ as the $1^{--}$ diquark-antidiquark type tetraquark state, more experimental data are still needed to distinguish its quark constituents.
There are no candidates for  the positive charge conjugation vector tetraquark states, the predictions can be confronted with the experimental data in the future at the BESIII, LHCb and Belle-II.
\end{abstract}



\section{Introduction}

 Recently, the BESIII collaboration  studied the process $e^+e^- \to (D^{*} \bar{D}^{*})^{\pm} \pi^\mp$ at
a center-of-mass energy of $4.26\,\rm{GeV}$ using a $827\,\rm{pb}^{-1}$ data sample
obtained with the BESIII detector at the Beijing Electron Positron Collider, and observed
 a structure $Z^{\pm}_c(4025)$ near the $(D^{*} \bar{D}^{*})^{\pm}$ threshold in the $\pi^\mp$ recoil mass spectrum \cite{BES1308}.
 The measured mass and width of the $Z^{\pm}_c(4025)$ are $(4026.3\pm2.6\pm3.7)\,\rm{MeV}$  and $(24.8\pm5.6\pm7.7)\,\rm{MeV}$, respectively \cite{BES1308}.
Later, the  BESIII collaboration studied the process $e^+e^- \to \pi^+\pi^- h_c$ at center-of-mass energies from $3.90\,\rm{ GeV}$ to $4.42\,\rm{GeV}$, and observed a distinct structure  $Z_c(4020)$   in the $\pi^\pm h_c$ mass spectrum,  the measured mass and width of the $Z_c(4020)$ are $(4022.9\pm 0.8\pm 2.7)\,\rm{MeV}$   and $(7.9\pm 2.7\pm 2.6)\,\rm{MeV}$, respectively \cite{BES1309}.  No significant signal of the $Z_c(3900)$ was observed in the $\pi^\pm h_c$ mass spectrum \cite{BES1309}, the $Z_c(3900)$ and $Z_c(4020)$ maybe have different quantum numbers.

At first sight, the S-wave $D^{*} \bar{D}^{*}$ systems  have the quantum numbers $J^{PC}=0^{++}$, $1^{+-}$, $2^{++}$, while the S-wave $ \pi^\pm h_c$ systems have the quantum numbers $J^{PC}=1^{--}$, so the $Z_c(4025)$ and $Z_c(4020)$ are different particles. On the other hand, it is also possible for the P-wave $D^{*} \bar{D}^{*}$ ($h_c\pi$) systems to have the quantum numbers $J^{PC}=1^{--}$ ($1^{+-}$). We cannot exclude the possibility that the $Z_c(4025)$ and $Z_c(4020)$ are the same particle with the quantum numbers $J^{PC}=1^{--}$ or $1^{+-}$. There have been several tentative assignments of the  $Z_c(4025)$ and $Z_c(4020)$, such as the re-scattering effects \cite{Rescatter}, molecular states \cite{Molecule}, tetraquark states \cite{Tetraquark-Qiao}, etc. The $Z_c(4025)$ and $Z_c(4020)$ are charged charmonium-like states, their quark constituents must be $c\bar{c}u\bar{d}$ or $c\bar{c}d\bar{u}$ irrespective of the diquark-antidiquark type or meson-meson type substructures.

In 2013, the BESIII collaboration studied  the process  $e^+e^- \to \pi^+\pi^-J/\psi$  and observed the $Z_c(3900)$ in the $\pi^\pm J/\psi$ mass spectrum with the mass $(3899.0\pm 3.6\pm 4.9)\,\rm{ MeV}$ and  width  $(46\pm 10\pm 20) \,\rm{MeV}$, respectively \cite{BES3900}. Later the $Z_c(3900)$ was confirmed by the Belle and CLEO collaborations \cite{Belle3900,CLEO3900}. Also in 2013, the BESIII collaboration studied  the process $e^+e^- \to \pi^{\mp} \left(D \bar{D}^*\right)^{\pm}$ and observed the  $Z_c(3885)$  in the $(D \bar{D}^*)^{\pm}$   mass spectrum with the mass  $(3883.9 \pm 1.5 \pm 4.2)\,\rm{ MeV}$ and width  $(24.8 \pm 3.3 \pm 11.0)\,\rm{ MeV}$, respectively  \cite{BES-3885}.   The angular distribution of the $\pi Z_c(3885)$ system favors assigning the $Z_c(3885)$ with $J^P=1^+$ \cite{BES-3885}. We tentatively  identify the $Z_c(3900)$
 and $Z_c(3885)$ as the same particle according to the uncertainties of the masses and widths \cite{WangHuangTao}, one can consult Ref.\cite{WangHuangTao} for more articles on the $Z_c(3900)$.
  The possible quantum numbers of the $Z_c(3900)$ or  $Z_c(3885)$ are $J^{PC}=1^{+-}$.   There is a faint possibility that the $Z_c(3900)$ and $Z_c(4020)$ are the same axial-vector meson with $J^{PC}=1^{+-}$ according to the masses.

In 2007, the  Belle collaboration  measured  the cross section for the process $e^+e^- \to \pi^+ \pi^- \psi^{\prime}$ between threshold and $\sqrt{s}=5.5\,\rm{GeV}$  using a $673\,\rm{fb}^{-1}$ data sample collected with the Belle detector at KEKB, and  observed two structures $Y(4360)$ and $Y(4660)$ in the $\pi^+ \pi^- \psi^{\prime}$ invariant mass distributions   at $(4361\pm 9\pm 9)\, \rm{MeV}$ with a width of \, $(74\pm 15\pm 10)\,\rm{ MeV}$ and   $(4664\pm 11\pm 5)\,\rm{ MeV}$ with a width of \, $(48\pm 15\pm 3) \,\rm{MeV}$, respectively \cite{Belle4660-0707}. The quantum numbers of the $Y(4360)$ and $Y(4660)$ are $J^{PC}=1^{--}$, which are unambiguously  listed in the  Review of Particle Physics now \cite{PDG}.
In 2008, the Belle collaboration studied  the exclusive process $e^+e^- \to \Lambda_c^+ \Lambda_c^-$   and observed a clear peak $Y(4630)$   in the $\Lambda_c^+ \Lambda_c^-$  invariant mass distribution just above the $\Lambda_c^+ \Lambda_c^-$ threshold, and determined  the mass and width to be   $\left(4634^{+8}_{-7}{}^{+5}_{-8}\right)\,\rm{Mev}$ and $\left(92^{+40}_{-24}{}^{+10}_{-21}\right)\,\rm{MeV}$, respectively \cite{Belle4630-0807}. The $Y(4660)$ and $Y(4630)$ may be the same particle according to the uncertainties of the masses and widths (also the decay properties \cite{Guo4660-4630}). There have been several tentative assignments of the $Y(4360)$ and $Y(4660)$, such as the conventional charmonium states \cite{DY4660}, baryonium state \cite{Qiao4660}, molecular states or hadro-charmonium states \cite{GuoWang}, tetraquark states \cite{Tetraquark-Neilsen-4660,Tetraquark-Ebert,Tetraquark-Ebert-2}, etc. One can consult Ref.\cite{Swanson2006} for more articles on the $X$, $Y$ and $Z$ particles.

In this article, we study the diquark-antidiquark type vector tetraquark states in details  with the QCD sum rules,  and explore possible assignments of the  $Z_c(4020)$, $Z_c(4025)$, $Y(4360)$ and $Y(4660)$ in the tetraquark scenario. In Ref.\cite{WangHuangTao},  we extend our previous works on the axial-vector tetraquark states \cite{Wang-Axial},  distinguish
the charge conjugations of the interpolating currents,  calculate the contributions of the vacuum condensates up to dimension-10 and discard the perturbative corrections in the operator product expansion, study the $C\gamma_5-C\gamma_\mu$ type axial-vector hidden charmed tetraquark states with the QCD sum rules. We explore the energy scale dependence of the charmed tetraquark states  in details for the first time, and   tentatively  assign the $X(3872)$ and $Z_c(3900)$ (or $Z_c(3885)$) as the $J^{PC}=1^{++}$ and $1^{+-}$ tetraquark states, respectively \cite{WangHuangTao}. In calculations, we observe that  the tetraquark  masses decrease monotonously with increase of the energy scales, the energy scale $\mu=1.5\,\rm{GeV}$ is the lowest energy scale to reproduce the experimental values of the masses of the   $X(3872)$ and $Z_c(3900)$, and serves as an  acceptable   energy  scale for the charmed mesons in the QCD sum rules \cite{WangHuangTao}.

In Ref.\cite{WangJPG}, we study the $C\gamma_\mu-C$ and $C\gamma_\mu\gamma_5-C\gamma_5$ type   tetraquark states with the QCD sum rules by  carrying  out the operator product expansion to the vacuum condensates   up to
dimension-10 and setting the energy scale to be $\mu=1\,\rm{GeV}$.
In Refs.\cite{Tetraquark-Qiao,Tetraquark-Neilsen-4660,Nielsen4430}, the authors carry   out the operator product expansion to the vacuum condensates  up to
dimension-8 to study the   vector tetraquark states with the QCD sum rules, but do not show  the energy scales or do not  specify the energy scales  at which the QCD spectral densities are calculated.
In Refs.\cite{Tetraquark-Qiao,Tetraquark-Neilsen-4660,WangJPG,Nielsen4430},  some higher dimension vacuum condensates involving the gluon condensate, mixed condensate and four-quark condensate are neglected, which maybe impair the predictive ability. The terms associate with $\frac{1}{T^2}$, $\frac{1}{T^4}$, $\frac{1}{T^6}$ in the QCD spectral densities  manifest themselves at small values of the Borel parameter $T^2$, we have to choose large values of the $T^2$ to warrant convergence of the operator product expansion and appearance of the Borel platforms. In the Borel windows, the higher dimension vacuum condensates  play a less important role.
In summary, the higher dimension vacuum condensates play an important role in determining the Borel windows therefore the ground state  masses and pole residues, so we should take them into account consistently.

In this article, we  extend our previous works \cite{WangHuangTao} to study the vector tetraquark states, distinguish
the charge conjugations of the interpolating currents,  calculate the contributions of the vacuum condensates up to dimension-10 and discard the perturbative corrections, study the masses and pole residues of the $C-C\gamma_\mu$ type vector hidden charmed tetraquark states with the QCD sum rules. Furthermore, we explore the energy scale dependence in details so as to obtain some useful formulae, and make tentative assignments  of the $Z_c(4020)$, $Z_c(4025)$, $Y(4360)$ and $Y(4660)$ as the $J^{PC}=1^{-+}$ or  $1^{--}$ tetraquark states. The scalar and axial-vector heavy-light diquark states have almost  degenerate masses from the QCD sum rules \cite{WangDiquark}, the $C\gamma_\mu-C$ and $C\gamma_\mu\gamma_5-C\gamma_5$ type   tetraquark states  have degenerate (or slightly different) masses \cite{WangJPG}, as  the pseudoscalar and vector heavy-light diquark states have slightly different masses.

The article is arranged as follows:  we derive the QCD sum rules for the masses and pole residues of  the vector tetraquark states  in section 2; in section 3, we present the numerical results and discussions; section 4 is reserved for our conclusion.

\section{QCD sum rules for  the  vector tetraquark states }
In the following, we write down  the two-point correlation functions $\Pi_{\mu\nu}(p)$  in the QCD sum rules,
\begin{eqnarray}
\Pi_{\mu\nu}(p)&=&i\int d^4x e^{ip \cdot x} \langle0|T\left\{J_\mu(x)J_\nu^{\dagger}(0)\right\}|0\rangle \, , \\
J^1_\mu(x)&=&\frac{\epsilon^{ijk}\epsilon^{imn}}{\sqrt{2}}\left\{s^j(x)C c^k(x) \bar{s}^m(x)\gamma_\mu C \bar{c}^n(x)+ts^j(x)C\gamma_\mu c^k(x)\bar{s}^m(x)C \bar{c}^n(x) \right\} \, , \\
J^2_\mu(x)&=&\frac{\epsilon^{ijk}\epsilon^{imn}}{2}\left\{u^j(x)C c^k(x) \bar{u}^m(x)\gamma_\mu C \bar{c}^n(x)+d^j(x)C c^k(x) \bar{d}^m(x)\gamma_\mu C \bar{c}^n(x) \right.\nonumber\\
&&\left.+tu^j(x)C\gamma_\mu c^k(x)\bar{u}^m(x)C \bar{c}^n(x) +td^j(x)C\gamma_\mu c^k(x)\bar{d}^m(x)C \bar{c}^n(x)\right\} \, , \\
 J^3_\mu(x)&=&\frac{\epsilon^{ijk}\epsilon^{imn}}{\sqrt{2}}\left\{u^j(x)C c^k(x) \bar{d}^m(x)\gamma_\mu C \bar{c}^n(x)+tu^j(x)C\gamma_\mu c^k(x)\bar{d}^m(x)C \bar{c}^n(x) \right\} \, ,
\end{eqnarray}
where $J_\mu(x)=J_\mu^1(x),\,J_\mu^2(x),\,J_\mu^3(x)$,  $t=\pm 1$, the $i$, $j$, $k$, $m$, $n$ are color indexes, the $C$ is the charge conjugation matrix.
 Under charge conjugation transform $\widehat{C}$, the currents $J_\mu(x)$ have the properties,
\begin{eqnarray}
\widehat{C}J^{1/2}_{\mu}(x)\widehat{C}^{-1}&=&\pm J^{1/2}_\mu(x) \,\,\,\, {\rm for}\,\,\,\, t=\pm1\, , \nonumber\\
\widehat{C}J^{3}_{\mu}(x)\widehat{C}^{-1}&=&\pm J^{3}_\mu(x)\mid_{u\leftrightarrow d} \,\,\,\, {\rm for}\,\,\,\, t=\pm1\, ,
\end{eqnarray}
which originate from the charge conjugation properties of the pseudoscalar and axial-vector diquark states,
\begin{eqnarray}
\widehat{C}\left[\epsilon^{ijk}q^j C  c^k\right]\widehat{C}^{-1}&=&\epsilon^{ijk}\bar{q}^j   C \bar{c}^k \, , \nonumber\\
\widehat{C}\left[\epsilon^{ijk}q^j C\gamma_\mu c^k\right]\widehat{C}^{-1}&=&\epsilon^{ijk}\bar{q}^j \gamma_\mu C \bar{c}^k \, .
\end{eqnarray}

We choose  the  neutral  currents $J^1_\mu(x)$ and $J^2_\mu(x)$ with $t=-$ to interpolate the  $J^{PC}=1^{--}$ diquark-antidiquark type tetraquark states $Y(4660)$ and $Y(4360)$, respectively. There are  two structures in $\pi^+\pi^-$ invariant mass distributions  at about $0.6\,\rm{GeV}$ and $1.0\,\rm{GeV}$ in the $\pi^+ \pi^-\psi^{\prime}$ mass spectrum, which maybe due to the scalar mesons $f_0(600)$ and $f_0(980)$, respectively \cite{Belle4660-0707}. In the two-quark scenario, $f_0(600)=(u\overline{u}+d\overline{d})/\sqrt{2}$ and $f_0=s\overline{s}$ in the ideal mixing limit, while in the tetraquark scenario, the $f_0(600)$ and $f_0(980)$ have the symbolic quark structures $ud\bar{u}\bar{d}$ and $(us\bar{u}\bar{s}+
ds\bar{d}\bar{s})/\sqrt{2}$, respectively.
The $Y(4660)$ couples    to the current $J^1_\mu(x)$ while the $Y(4360)$ couples   to the current $J^2_\mu(x)$. However, we cannot exclude the possibility that the $Y(4660)$ has the  symbolic quark structure $c\bar{c}(u\bar{u}+d\bar{d})/\sqrt{2}$, in that case the decay $Y(4660)\to f_0(600)\psi^{\prime}$ is Okubo-Zweig-Iizuka (OZI) allowed.  We choose the charged vector current $J^3_\mu(x)$ with $t=\pm$ to interpolate the $Z_c(4020)$ and $Z_c(4025)$, the results for the scalar and tensor currents will be presented elsewhere. At present time, we cannot exclude the possibility that the $Z_c(4020)$ and $Z_c(4025)$ are the same vector particle.

We can insert  a complete set of intermediate hadronic states with
the same quantum numbers as the current operators $J_\mu(x)$ into the
correlation functions $\Pi_{\mu\nu}(p)$  to obtain the hadronic representation
\cite{SVZ79,Reinders85}. After isolating the ground state
contributions of the vector tetraquark states, we get the following results,
\begin{eqnarray}
\Pi_{\mu\nu}(p)&=&\frac{\lambda_{Y/Z}^2}{M_{Y/Z}^2-p^2}\left(-g_{\mu\nu} +\frac{p_\mu p_\nu}{p^2}\right) +\cdots \, \, ,
\end{eqnarray}
where the pole residues  $\lambda_{Y/Z}$ are defined by
\begin{eqnarray}
 \langle 0|J_\mu(0)|Y/Z(p)\rangle=\lambda_{Y/Z} \,\varepsilon_\mu \, ,
\end{eqnarray}
the $\varepsilon_\mu$ are the polarization vectors of the  vector tetraquark states $Z_c(4020)$, $Z_c(4025)$, $Y(4360)$, $Y(4660)$, etc.

 In the following,  we take the current $J_\mu(x)=J^1_\mu(x)$ as an example and briefly outline  the operator product expansion for the correlation functions $\Pi_{\mu\nu}(p)$  in perturbative QCD.  We contract the $c$ and $s$ quark fields in the correlation functions
$\Pi_{\mu\nu}(p)$ with Wick theorem, obtain the results:
\begin{eqnarray}
\Pi_{\mu\nu}(p)&=&\frac{i\epsilon^{ijk}\epsilon^{imn}\epsilon^{i^{\prime}j^{\prime}k^{\prime}}\epsilon^{i^{\prime}m^{\prime}n^{\prime}}}{2}\int d^4x e^{ip \cdot x}   \nonumber\\
&&\left\{{\rm Tr}\left[ C^{kk^{\prime}}(x) CS^{jj^{\prime}T}(x)C\right] {\rm Tr}\left[ \gamma_\nu C^{n^{\prime}n}(-x)\gamma_\mu C S^{m^{\prime}mT}(-x)C\right] \right. \nonumber\\
&&+{\rm Tr}\left[ \gamma_\mu C^{kk^{\prime}}(x)\gamma_\nu CS^{jj^{\prime}T}(x)C\right] {\rm Tr}\left[  C^{n^{\prime}n}(-x) C S^{m^{\prime}mT}(-x)C\right] \nonumber\\
&&\mp{\rm Tr}\left[ \gamma_\mu C^{kk^{\prime}}(x) CS^{jj^{\prime}T}(x)C\right] {\rm Tr}\left[ \gamma_\nu C^{n^{\prime}n}(-x) C S^{m^{\prime}mT}(-x)C\right] \nonumber\\
 &&\left.\mp{\rm Tr}\left[  C^{kk^{\prime}}(x)\gamma_\nu CS^{jj^{\prime}T}(x)C\right] {\rm Tr}\left[  C^{n^{\prime}n}(-x)\gamma_\mu C S^{m^{\prime}mT}(-x)C\right] \right\} \, ,
\end{eqnarray}
where the $\mp$ correspond to $C=\pm$ respectively,
 the $S_{ij}(x)$ and $C_{ij}(x)$ are the full $s$ and $c$ quark propagators respectively,
 \begin{eqnarray}
S_{ij}(x)&=& \frac{i\delta_{ij}\!\not\!{x}}{ 2\pi^2x^4}
-\frac{\delta_{ij}m_s}{4\pi^2x^2}-\frac{\delta_{ij}\langle
\bar{s}s\rangle}{12} +\frac{i\delta_{ij}\!\not\!{x}m_s
\langle\bar{s}s\rangle}{48}-\frac{\delta_{ij}x^2\langle \bar{s}g_s\sigma Gs\rangle}{192}+\frac{i\delta_{ij}x^2\!\not\!{x} m_s\langle \bar{s}g_s\sigma
 Gs\rangle }{1152}\nonumber\\
&& -\frac{ig_s G^{a}_{\alpha\beta}t^a_{ij}(\!\not\!{x}
\sigma^{\alpha\beta}+\sigma^{\alpha\beta} \!\not\!{x})}{32\pi^2x^2} -\frac{i\delta_{ij}x^2\!\not\!{x}g_s^2\langle \bar{s} s\rangle^2}{7776} -\frac{\delta_{ij}x^4\langle \bar{s}s \rangle\langle g_s^2 GG\rangle}{27648}-\frac{1}{8}\langle\bar{s}_j\sigma^{\mu\nu}s_i \rangle \sigma_{\mu\nu} \nonumber\\
&&   -\frac{1}{4}\langle\bar{s}_j\gamma^{\mu}s_i\rangle \gamma_{\mu }+\cdots \, ,
\end{eqnarray}
\begin{eqnarray}
C_{ij}(x)&=&\frac{i}{(2\pi)^4}\int d^4k e^{-ik \cdot x} \left\{
\frac{\delta_{ij}}{\!\not\!{k}-m_c}
-\frac{g_sG^n_{\alpha\beta}t^n_{ij}}{4}\frac{\sigma^{\alpha\beta}(\!\not\!{k}+m_c)+(\!\not\!{k}+m_c)
\sigma^{\alpha\beta}}{(k^2-m_c^2)^2}\right.\nonumber\\
&&\left. +\frac{g_s D_\alpha G^n_{\beta\lambda}t^n_{ij}(f^{\lambda\beta\alpha}+f^{\lambda\alpha\beta}) }{3(k^2-m_c^2)^4}-\frac{g_s^2 (t^at^b)_{ij} G^a_{\alpha\beta}G^b_{\mu\nu}(f^{\alpha\beta\mu\nu}+f^{\alpha\mu\beta\nu}+f^{\alpha\mu\nu\beta}) }{4(k^2-m_c^2)^5}+\cdots\right\} \, ,\nonumber\\
f^{\lambda\alpha\beta}&=&(\!\not\!{k}+m_c)\gamma^\lambda(\!\not\!{k}+m_c)\gamma^\alpha(\!\not\!{k}+m_c)\gamma^\beta(\!\not\!{k}+m_c)\, ,\nonumber\\
f^{\alpha\beta\mu\nu}&=&(\!\not\!{k}+m_c)\gamma^\alpha(\!\not\!{k}+m_c)\gamma^\beta(\!\not\!{k}+m_c)\gamma^\mu(\!\not\!{k}+m_c)\gamma^\nu(\!\not\!{k}+m_c)\, ,
\end{eqnarray}
and  $t^n=\frac{\lambda^n}{2}$, the $\lambda^n$ is the Gell-Mann matrix,  $D_\alpha=\partial_\alpha-ig_sG^n_\alpha t^n$ \cite{Reinders85}, then compute  the integrals both in the coordinate and momentum spaces,  and obtain the correlation functions $\Pi_{\mu\nu}(p)$ therefore the spectral densities at the level of   quark-gluon degrees  of freedom. In Eq.(10), we retain the terms $\langle\bar{s}_j\sigma_{\mu\nu}s_i \rangle$ and $\langle\bar{s}_j\gamma_{\mu}s_i\rangle$ originate from the Fierz re-arrangement of the $\langle s_i \bar{s}_j\rangle$ to  absorb the gluons  emitted from the heavy quark lines to form
$\langle\bar{s}_j g_s G^a_{\alpha\beta} t^a_{mn}\sigma_{\mu\nu} s_i \rangle$ and $\langle\bar{s}_j\gamma_{\mu}s_ig_s D_\nu G^a_{\alpha\beta}t^a_{mn}\rangle$ so as to extract the mixed condensate and four-quark condensates $\langle\bar{s}g_s\sigma G s\rangle$ and $g_s^2\langle\bar{s}s\rangle^2$, respectively.
One can consult Ref.\cite{WangHuangTao} for some technical details in the operator product expansion.

 Once analytical results are obtained,  we can take the
quark-hadron duality below the continuum threshold $s_0$ and perform Borel transform  with respect to
the variable $P^2=-p^2$ to obtain  the following QCD sum rules:
\begin{eqnarray}
\lambda^2_{Y/Z}\, \exp\left(-\frac{M^2_{Y/Z}}{T^2}\right)= \int_{4m_c^2}^{s_0} ds\, \rho(s) \, \exp\left(-\frac{s}{T^2}\right) \, ,
\end{eqnarray}
where
\begin{eqnarray}
\rho(s)&=&\rho_{0}(s)+\rho_{3}(s) +\rho_{4}(s)+\rho_{5}(s)+\rho_{6}(s)+\rho_{7}(s) +\rho_{8}(s)+\rho_{10}(s)\, ,
\end{eqnarray}
 the 0, 3, 4, 5, 6, 7, 8, 10 denote the dimensions of the vacuum condensates, the explicit expressions of the  spectral densities $\rho_i(s)$ are presented in the Appendix.
 In this article, we carry out the
operator product expansion to the vacuum condensates  up to dimension-10 and discard the  perturbative corrections, and
assume vacuum saturation for the  higher dimension vacuum condensates. The higher
dimension vacuum condensates  are always
 factorized to lower condensates with vacuum saturation in the QCD sum rules,
  factorization works well in  large $N_c$ limit.  In reality, $N_c=3$, some  (not much) ambiguities maybe come from
the vacuum saturation assumption.
The condensates $\langle \frac{\alpha_s}{\pi}GG\rangle$, $\langle \bar{s}s\rangle\langle \frac{\alpha_s}{\pi}GG\rangle$,
$\langle \bar{s}s\rangle^2\langle \frac{\alpha_s}{\pi}GG\rangle$, $\langle \bar{s} g_s \sigma Gs\rangle^2$ and $g_s^2\langle \bar{s}s\rangle^2$ are the vacuum expectations
of the operators of the order
$\mathcal{O}(\alpha_s)$.  The four-quark condensate $g_s^2\langle \bar{s}s\rangle^2$ comes from the terms
$\langle \bar{s}\gamma_\mu t^a s g_s D_\eta G^a_{\lambda\tau}\rangle$, $\langle\bar{s}_jD^{\dagger}_{\mu}D^{\dagger}_{\nu}D^{\dagger}_{\alpha}s_i\rangle$  and
$\langle\bar{s}_jD_{\mu}D_{\nu}D_{\alpha}s_i\rangle$, rather than comes from the perturbative corrections of $\langle \bar{s}s\rangle^2$ (see Ref.\cite{WangHuangTao} for the technical details).
 The condensates $\langle g_s^3 GGG\rangle$, $\langle \frac{\alpha_s GG}{\pi}\rangle^2$,
 $\langle \frac{\alpha_s GG}{\pi}\rangle\langle \bar{s} g_s \sigma Gs\rangle$ have the dimensions 6, 8, 9 respectively,  but they are   the vacuum expectations
of the operators of the order    $\mathcal{O}( \alpha_s^{3/2})$, $\mathcal{O}(\alpha_s^2)$, $\mathcal{O}( \alpha_s^{3/2})$ respectively, and discarded.  We take
the truncations $n\leq 10$ and $k\leq 1$ in a consistent way,
the operators of the orders $\mathcal{O}( \alpha_s^{k})$ with $k> 1$ are  discarded. Furthermore,  the numerical values of the  condensates $\langle g_s^3 GGG\rangle$, $\langle \frac{\alpha_s GG}{\pi}\rangle^2$,
 $\langle \frac{\alpha_s GG}{\pi}\rangle\langle \bar{s} g_s \sigma Gs\rangle$   are very small, and  they are neglected safely.

 Differentiate   Eq.(12) with respect to  $\frac{1}{T^2}$, then eliminate the
 pole residues $\lambda_{Y/Z}$, we obtain the QCD sum rules for
 the masses of the vector   tetraquark states,
 \begin{eqnarray}
 M^2_{Y/Z}= \frac{\int_{4m_c^2}^{s_0} ds\frac{d}{d \left(-1/T^2\right)}\rho(s)\exp\left(-\frac{s}{T^2}\right)}{\int_{4m_c^2}^{s_0} ds \rho(s)\exp\left(-\frac{s}{T^2}\right)}\, .
\end{eqnarray}

We can obtain the QCD sum rules for the vector tetraquark states $c\bar{c}u\bar{d}$ and   $c\bar{c}(u\bar{u}+d\bar{d})/\sqrt{2}$ with the simple replacements,
 \begin{eqnarray}
m_s &\to& 0\, , \nonumber\\
\langle\bar{s}s\rangle &\to& \langle\bar{q}q\rangle\, , \nonumber\\
\langle\bar{s}g_s\sigma G s\rangle &\to& \langle\bar{q}g_s\sigma Gq\rangle\, ,
\end{eqnarray}
 the QCD sum rules for the $c\bar{c}u\bar{d}$ and   $c\bar{c}(u\bar{u}+d\bar{d})/\sqrt{2}$  degenerate in the isospin limit.

\section{Numerical results and discussions}
The vacuum condensates are taken to be the standard values
$\langle\bar{q}q \rangle=-(0.24\pm 0.01\, \rm{GeV})^3$,  $\langle\bar{s}s \rangle=(0.8\pm0.1)\langle\bar{q}q \rangle$,
$\langle\bar{q}g_s\sigma G q \rangle=m_0^2\langle \bar{q}q \rangle$, $\langle\bar{s}g_s\sigma G s \rangle=m_0^2\langle \bar{s}s \rangle$,
$m_0^2=(0.8 \pm 0.1)\,\rm{GeV}^2$, $\langle \frac{\alpha_s
GG}{\pi}\rangle=(0.33\,\rm{GeV})^4 $    at the energy scale  $\mu=1\, \rm{GeV}$
\cite{SVZ79,Reinders85,Ioffe2005}.
The quark condensate and mixed quark condensate evolve with the   renormalization group equation,
$\langle\bar{q}q \rangle(\mu)=\langle\bar{q}q \rangle(Q)\left[\frac{\alpha_{s}(Q)}{\alpha_{s}(\mu)}\right]^{\frac{4}{9}}$,
 $\langle\bar{s}s \rangle(\mu)=\langle\bar{s}s \rangle(Q)\left[\frac{\alpha_{s}(Q)}{\alpha_{s}(\mu)}\right]^{\frac{4}{9}}$,
 $\langle\bar{q}g_s \sigma Gq \rangle(\mu)=\langle\bar{q}g_s \sigma Gq \rangle(Q)\left[\frac{\alpha_{s}(Q)}{\alpha_{s}(\mu)}\right]^{\frac{2}{27}}$
 and $\langle\bar{s}g_s \sigma Gs \rangle(\mu)=\langle\bar{s}g_s \sigma Gs \rangle(Q)\left[\frac{\alpha_{s}(Q)}{\alpha_{s}(\mu)}\right]^{\frac{2}{27}}$.

In the article, we take the $\overline{MS}$ masses $m_{c}(m_c)=(1.275\pm0.025)\,\rm{GeV}$ and $m_s(\mu=2\,\rm{GeV})=(0.095\pm0.005)\,\rm{GeV}$
 from the Particle Data Group \cite{PDG}, and take into account
the energy-scale dependence of  the $\overline{MS}$ masses from the renormalization group equation,
\begin{eqnarray}
m_c(\mu)&=&m_c(m_c)\left[\frac{\alpha_{s}(\mu)}{\alpha_{s}(m_c)}\right]^{\frac{12}{25}} \, ,\nonumber\\
m_s(\mu)&=&m_s({\rm 2GeV} )\left[\frac{\alpha_{s}(\mu)}{\alpha_{s}({\rm 2GeV})}\right]^{\frac{4}{9}} \, ,\nonumber\\
\alpha_s(\mu)&=&\frac{1}{b_0t}\left[1-\frac{b_1}{b_0^2}\frac{\log t}{t} +\frac{b_1^2(\log^2{t}-\log{t}-1)+b_0b_2}{b_0^4t^2}\right]\, ,
\end{eqnarray}
  where $t=\log \frac{\mu^2}{\Lambda^2}$, $b_0=\frac{33-2n_f}{12\pi}$, $b_1=\frac{153-19n_f}{24\pi^2}$, $b_2=\frac{2857-\frac{5033}{9}n_f+\frac{325}{27}n_f^2}{128\pi^3}$,  $\Lambda=213\,\rm{MeV}$, $296\,\rm{MeV}$  and  $339\,\rm{MeV}$ for the flavors  $n_f=5$, $4$ and $3$, respectively  \cite{PDG}.

\begin{figure}
\centering
\includegraphics[totalheight=6cm,width=7cm]{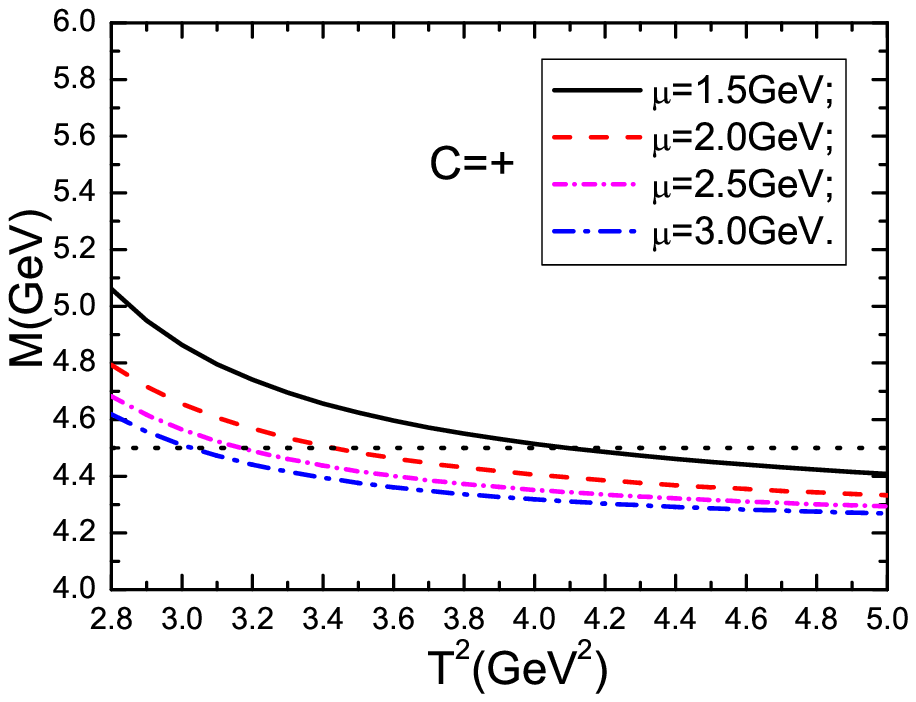}
\includegraphics[totalheight=6cm,width=7cm]{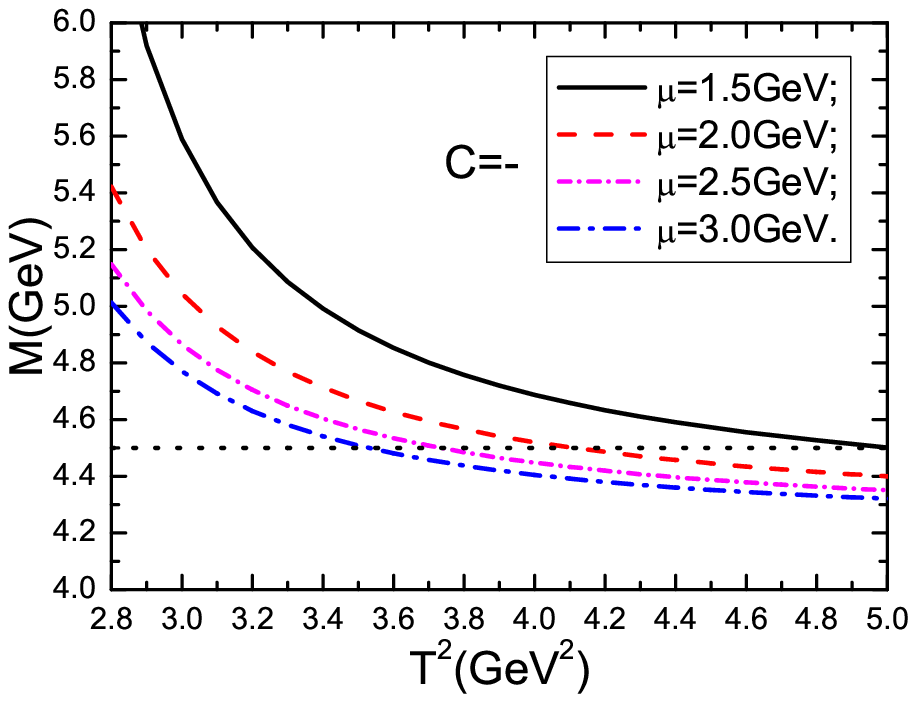}
\includegraphics[totalheight=6cm,width=7cm]{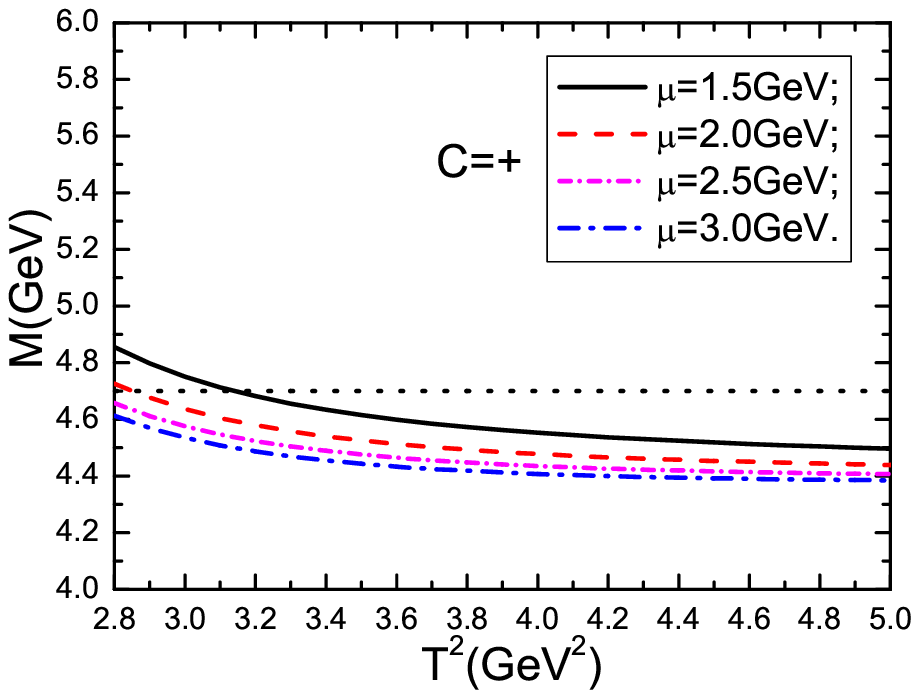}
\includegraphics[totalheight=6cm,width=7cm]{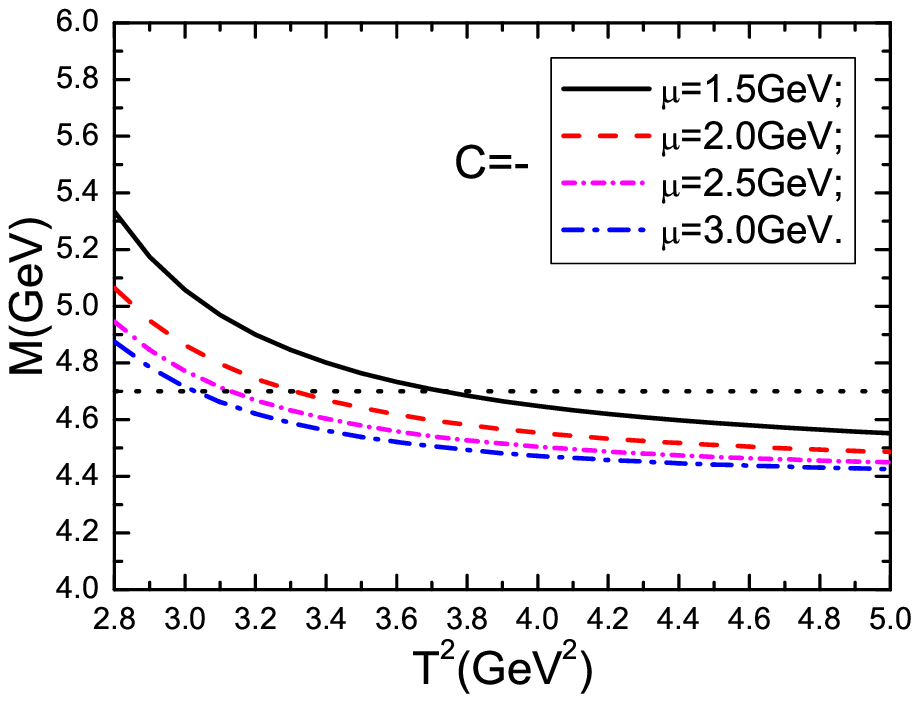}
  \caption{ The masses of the vector $c\bar{c}u\bar{d}$ tetraquark states with variations of the  Borel parameters $T^2$, energy scales $\mu$ and threshold parameters $\sqrt{s_0}$, where the horizontal lines denote the threshold parameters $\sqrt{s_0}=4.5\,\rm{GeV}$ and $4.7\,\rm{GeV}$, respectively; the $C=\pm$ denote the charge conjugations. }
\end{figure}

In Ref.\cite{WangHuangTao}, we observe that  the energy scale  $\mu=(1.1-1.6)\,\rm{GeV}$ is an acceptable   energy scale of the QCD spectral densities in the QCD sum rules for the hidden and open charmed mesons, as it can reproduce the experimental values $M_D=1.87\,\rm{GeV}$ and $M_{J/\psi}=3.1\,\rm{GeV}$ with suitable   Borel parameters.
However, such energy scale  and truncation in the operator product expansion cannot reproduce the experimental values of the decay constants $f_{D}$ and $f_{J/\psi}$. In calculation, we observe that the masses of the axial-vector tetraquark states decrease monotonously with increase of the energy scales of the QCD spectral densities, the energy scale $\mu=1.5\,\rm{GeV}$ is the lowest energy scale to reproduce the experimental values  of the masses of the $X(3872)$ and $Z_c(3900)$ (or $Z_c(3885)$), and serves  as an acceptable energy scale  (not the universal energy scale) for the tetraquark  states \cite{WangHuangTao}.
On the other hand, it is hard to obtain the true values of the pole residues $\lambda_{X/Y/Z}$ of the tetraquark states, so we focus on the masses to study the tetraquark states, and the predictions of the pole
residues maybe not as robust. If the $Z_c(4020)$ and $Z_c(4025)$ are the vector tetraquark states, we can choose the threshold parameters   $\sqrt{s_0}=(4.3-4.8)\,\rm{GeV}$ and energy scales  $\mu=(1.5-3.0)\,\rm{GeV}$  tentatively, and search  for the ideal parameters, such as the threshold parameters, energy scales and Borel parameters.

In Fig.1,  the masses of the vector $c\bar{c}u\bar{d}$ tetraquark states are plotted   with variations of the  Borel parameters $T^2$, energy scales $\mu$, and continuum  threshold parameters $\sqrt{s_0}$. From the figure, we can see that the masses decrease monotonously with increase of the energy scales, the parameters $\sqrt{s_0}\leq 4.5\,\rm{GeV}$ and $\mu \leq 1.5\,\rm{GeV}$ can be excluded, as the predicted masses $M_{Z}\gg ({\rm or\,>})\sqrt{s_0}=4.5\,\rm{GeV}$ for the values of the  Borel parameters at a large interval. We have to choose larger threshold parameters or (and) energy scales, the resulting  masses are larger than  $4.3\,\rm{GeV}$ for the parameters $\sqrt{s_0}\geq4.5\,\rm{GeV}$ and $\mu=3.0\,\rm{GeV}$. The predictions based on the QCD sum rules disfavor  assigning the $Z_c(4020)$ and $Z_c(4025)$ as the  diquark-antidiquark type vector tetraquark states. We cannot satisfy the relation $\sqrt{s_0}=M_Z+0.5\,\rm{GeV}$ with reasonable  $M_Z$ compared to the experimental data.

The BESIII collaboration  observed the  $Z^{\pm}_c(4025)$ and $Z^{\pm}_c(4020)$ in the following processes \cite{BES1308,BES1309},
\begin{eqnarray}
e^+ e^- &\to&Z^{\pm}_c(4025)\pi^{\mp} \to (D^*\bar{D}^*)^{\pm}(0^{++},1^{+-},2^{++},0^{-+},1^{--},2^{-+},3^{--})\,\pi^{\mp} \, , \nonumber\\
e^+ e^- &\to&Z^{\pm}_c(4020)\pi^{\mp} \to (h_c\pi)^{\pm}(1^{--},0^{++},1^{+-},2^{++})\,\pi^{\mp} \, ,
\end{eqnarray}
where we present the possible quantum numbers $J^{PC}$ of the  $(D^*\bar{D}^*)^{\pm}$ and $(h_c\pi)^{\pm}$ systems in the brackets.
If the $Z^{\pm}_c(4025)$ and $Z^{\pm}_c(4020)$ are the same particle, the quantum numbers are $J^{PC}=1^{--}$, $0^{++}$, $1^{+-}$, $2^{++}$. On the other hand, the
$Z^{\pm}_c(4025)\pi^{\mp}$ and $Z^{\pm}_c(4020)\pi^{\mp}$ systems have the quantum numbers $J^{PC}=1^{--}$,  then the  survived quantum numbers of  the $Z^{\pm}_c(4025)$ and $Z^{\pm}_c(4020)$ are $J^{PC}=1^{--}$, $1^{+-}$ and $2^{++}$. The predictions  based on the QCD sum rules reduce the possible quantum numbers of the  $Z_c(4025)$ and $Z_c(4020)$  to $J^{PC}=1^{+-}$ and $2^{++}$.

The strong decays
\begin{eqnarray}
Y(4260)/\gamma^*(4260) &\to& Z_c^{\pm}(4025/4020)(2^{++})\,\pi^{\mp}\, ,
\end{eqnarray}
 take place through relative D-wave, and are  kinematically suppressed in the phase-space. The $2^{++}$ assignment is disfavored, but not excluded.

In the following, we list out the possible strong decays of the $Z^{\pm}_c(4025)$ and $Z^{\pm}_c(3900)$ in the case of the $J^{PC}=1^{+-}$ assignment.
\begin{eqnarray}
Z^{\pm}_c(4025)(1^{+-}) &\to& h_c({\rm 1P})\pi^{\pm}\, , \, J/\psi\pi^{\pm}\, , \, \eta_c \rho^{\pm}\, , \, \eta_c(\pi\pi)_{\rm P}^{\pm}
\, , \,\chi_{c1}(\pi\pi)_{\rm P}^{\pm}\, , \,(D\bar{D}^*)^{\pm}\, , \,(D^*\bar{D}^*)^{\pm}\, , \nonumber\\
Z^{\pm}_c(3900)(1^{+-}) &\to& h_c({\rm 1P})\pi^{\pm}\, , \, J/\psi\pi^{\pm}\, , \, \eta_c \rho^{\pm} \, , \, \eta_c(\pi\pi)_{\rm P}^{\pm} \, , \,\chi_{c1}(\pi\pi)_{\rm P}^{\pm}\, ,
\end{eqnarray}
where the  $(\pi\pi)_{\rm P}$ denotes  the P-wave $\pi\pi$ systems have the same quantum numbers of the $\rho$.  We take  the $Z_c(4025)$ and $Z_c(4020)$ as the same particle in the $J^{PC}=1^{+-}$ assignment, and will denote them as $Z_c(4025)$.
 In Ref.\cite{WangHuangTao}, we observe that the $Z_c(3900)$ couples   to the axial-vector current $J_{1^{+-}}^{\mu}$.
Now we perform Fierz re-arrangement  both in the color and Dirac-spinor  spaces and obtain the following result,
\begin{eqnarray}
J_{1^{+-}}^{\mu}&=&\frac{\epsilon^{ijk}\epsilon^{imn}}{\sqrt{2}}\left\{u^jC\gamma_5 c^k \bar{d}^m\gamma^\mu C \bar{c}^n-u^jC\gamma^\mu c^k\bar{d}^m\gamma_5 C \bar{c}^n \right\} \, , \nonumber\\
 &=&\frac{1}{2\sqrt{2}}\left\{\,i\bar{c}i\gamma_5 c\,\bar{d}\gamma^\mu u-i\bar{c} \gamma^\mu c\,\bar{d}i\gamma_5 u+\bar{c} u\,\bar{d}\gamma^\mu\gamma_5 c-\bar{c} \gamma^\mu \gamma_5u\,\bar{d}c\right. \nonumber\\
&&\left. - i\bar{c}\gamma_\nu\gamma_5c\, \bar{d}\sigma^{\mu\nu}u+i\bar{c}\sigma^{\mu\nu}c\, \bar{d}\gamma_\nu\gamma_5u
- i \bar{c}\sigma^{\mu\nu}\gamma_5u\,\bar{d}\gamma_\nu c+i\bar{c}\gamma_\nu u\, \bar{d}\sigma^{\mu\nu}\gamma_5c   \,\right\} \, ,
\end{eqnarray}
the components such as $\bar{c}i\gamma_5 c\,\bar{d}\gamma^\mu u$, $\bar{c} \gamma^\mu c\,\bar{d}i\gamma_5 u$, etc couple   to the  meson-meson  pairs,
the strong decays
\begin{eqnarray}
Z^{\pm}_c(3900)(1^{+-}) &\to& h_c({\rm 1P})\pi^{\pm}\, , \, J/\psi\pi^{\pm}\, , \, \eta_c \rho^{\pm} \, , \, \eta_c(\pi\pi)_{\rm P}^{\pm} \, ,
\end{eqnarray}
are OZI  super-allowed, we take the decays to the $(\pi\pi)_{\rm P}^{\pm}$   final states as OZI super-allowed according to the decays $\rho \to \pi\pi$.
The BESIII collaboration observed no evidence of the $Z_c(3900)$ in the process $e^+e^- \to \pi^+\pi^- h_c$ at center-of-mass energies from $3.90\,\rm{ GeV}$ to $4.42\,\rm{GeV}$  \cite{BES1309}. We expect to observe the $Z_c^{\pm}(3900)$ in the $h_c({\rm 1P})\pi^{\pm}$ final states  when a large amount of events are accumulated.
The $Z_c(4025)$  and $Z_c(3900)$ have the same quantum numbers and analogous  strong decays but different masses and quark configurations.

Now we take a short digression to discuss the interpolating currents consist of four quarks. The diquark-antidiquark type current with special quantum numbers couples  to a special tetraquark state, while the current can be re-arranged both in the color and Dirac-spinor  spaces, and changed  to a current as a special superposition of   color  singlet-singlet type currents.   The color  singlet-singlet type currents couple  to the meson-meson pairs. The
diquark-antidiquark type tetraquark state can be taken as a special superposition of a series of  meson-meson pairs, and embodies  the net effects. The decays to its components (meson-meson pairs) are OZI super-allowed,  the kinematically allowed decays take place easily.

We can search for the $Z^{\pm}_c(4025)(1^{+-})$ in the final states
$ h_c({\rm 1P})\pi^{\pm}$,  $J/\psi\pi^{\pm}$, $\eta_c \rho^{\pm}$, $\eta_c(\pi\pi)_{\rm P}^{\pm}$, $\chi_{c1}(\pi\pi)_{\rm P}^{\pm}$.
In Ref.\cite{Wangzg1312}, we observe that the $Z_c(4025)$ couples   to the axial-vector current $J_{1^{+-}}^{\mu\nu}$.
We perform Fierz re-arrangement both in the color and Dirac-spinor  spaces and obtain the following result,
\begin{eqnarray}
J_{1^{+-}}^{\mu\nu}&=&\frac{\epsilon^{ijk}\epsilon^{imn}}{\sqrt{2}}\left\{u^jC\gamma^\mu c^k \bar{d}^m\gamma^\nu C \bar{c}^n-u^jC\gamma^\nu c^k\bar{d}^m\gamma^\mu C \bar{c}^n \right\} \, , \nonumber\\
 &=&\frac{1}{2\sqrt{2}}\left\{\,i\bar{d}u\, \bar{c}\sigma^{\mu\nu}c +i\bar{d}\sigma^{\mu\nu}u \,\bar{c}c+i\bar{d}c\, \bar{c}\sigma^{\mu\nu}u +i\bar{d}\sigma^{\mu\nu}c \,\bar{c}u \right. \nonumber\\
 &&-\bar{c}\sigma^{\mu\nu}\gamma_5c\,\bar{d}i\gamma_5u-\bar{c}i\gamma_5 c\,\bar{d}\sigma^{\mu\nu}\gamma_5u -\bar{c}\sigma^{\mu\nu}\gamma_5u\,\bar{d}i\gamma_5c-\bar{d}i\gamma_5 c\,\bar{c}\sigma^{\mu\nu}\gamma_5u\nonumber\\
 &&+i\epsilon^{\mu\nu\alpha\beta}\bar{c}\gamma^\alpha\gamma_5c\, \bar{d}\gamma^\beta u-i\epsilon^{\mu\nu\alpha\beta}\bar{c}\gamma^\alpha c\, \bar{d}\gamma^\beta \gamma_5u\nonumber\\
 &&\left.+i\epsilon^{\mu\nu\alpha\beta}\bar{c}\gamma^\alpha\gamma_5u\, \bar{d}\gamma^\beta c-i\epsilon^{\mu\nu\alpha\beta}\bar{c}\gamma^\alpha u\, \bar{d}\gamma^\beta \gamma_5c \,\right\} \, .
\end{eqnarray}
 The scattering states $J/\psi\pi^{+}$, $\eta_c \rho^{+}$, $\eta_c(\pi\pi)_{\rm P}^{+}$,
$\chi_{c1}(\pi\pi)_{\rm P}^{+}$, $(DD^*)^+$ couple   to the components   $\bar{c}\sigma^{\mu\nu}\gamma_5c\,\bar{d}i\gamma_5u$,
$\bar{c}i\gamma_5 c\,\bar{d}\sigma^{\mu\nu}\gamma_5u $, $\bar{c}i\gamma_5 c\,\bar{d}\sigma^{\mu\nu}\gamma_5u $, $\epsilon^{\mu\nu\alpha\beta}\bar{c}\gamma^\alpha\gamma_5c\, \bar{d}\gamma^\beta u$, $\bar{c}\sigma^{\mu\nu}\gamma_5u\,\bar{d}i\gamma_5c$, respectively.
The strong decays
\begin{eqnarray}
Z^{\pm}_c(4025)(1^{+-}) &\to&  J/\psi\pi^{\pm}\, , \, \eta_c \rho^{\pm}\, , \, \eta_c(\pi\pi)_{\rm P}^{\pm}\, , \, \chi_{c1}(\pi\pi)_{\rm P}^{\pm} \, , \, (DD^*)^\pm \, ,
\end{eqnarray}
are OZI super-allowed. In this article, we take the decays to the $(\pi\pi)_{\rm P}^{\pm}/(\pi\pi\pi)_{\rm P}^{0}$   final states as OZI super-allowed according to the decays $\rho \to \pi\pi/\omega \to \pi\pi\pi$.

We can also search for the neutral partner $Z^{0}_c(4025)(1^{+-})$
in the following strong and electromagnetic decays,
\begin{eqnarray}
Z^{0}_c(4025)(1^{+-}) &\to& h_c({\rm 1P})\pi^{0} \, , \, J/\psi\pi^{0}\, , \, J/\psi \eta\, , \,\eta_c \rho^{0}\, , \, \eta_c \omega\, , \, \eta_c(\pi\pi)_{\rm P}^{0} \, , \, \chi_{cj}(\pi\pi)_{\rm P}^{0} \, , \,  \nonumber\\
&&\eta_c(\pi\pi\pi)_{\rm P}^{0} \, , \, \chi_{cj}(\pi\pi\pi)_{\rm P}^{0}\, , \,\eta_c \gamma \, , \, \chi_{cj} \gamma\, , \, (DD^*)^0\, ,
\end{eqnarray}
where the $(\pi\pi\pi)_{\rm P}$ denotes the P-wave $\pi\pi\pi$ systems with the same quantum numbers of the $\omega$.

On the other hand, if the $Z_c(4025)$ and $Z_c(4020)$ are different particles, we can search for the $Z^{\pm}_c(4025/4020)(0^{++})$ and $Z^{\pm}_c(4025/4020)(2^{++})$ in the following strong decays,
\begin{eqnarray}
Z^{\pm}_c(4025/4020)(0^{++}) &\to& \eta_c\pi^{\pm}\, , \, J/\psi\rho^{\pm}\, , \,  J/\psi(\pi\pi)_{\rm P}^{\pm}\, , \,\chi_{c1}\pi^{\pm}
\, ,\, D\bar{D}\, ,\, D^*\bar{D}^*\, , \nonumber\\
Z^{\pm}_c(4025/4020)(2^{++}) &\to& \eta_c\pi^{\pm}\, , \, J/\psi\rho^{\pm}\, , \, J/\psi(\pi\pi)_{\rm P}^{\pm}\, , \,\chi_{c1}\pi^{\pm}\, ,\, D\bar{D}\, ,\, D^*\bar{D}^*\, .
\end{eqnarray}
The strong decays
\begin{eqnarray}
Y(4260)/\gamma^*(4260) &\to& Z_c^{\pm}(4025/4020)(0^{++})\,\pi^{\mp}\, ,
\end{eqnarray}
cannot take place. The $0^{++}$ assignment is excluded.

Now, we explore the possibility of assigning  the $Y(4360)$ and $Y(4660)$ as the  diquark-antidiquark type vector tetraquark states.
 We consult the often used energy scale $\mu=\sqrt{m_D^2-m_c^2}\approx 1\,\rm{GeV}$ in the QCD sum rules for the $D$ mesons, and suggest  a  formula to estimate the energy scales of the QCD spectral densities in the QCD sum rules for the  hidden charmed tetraquark states,
 \begin{eqnarray}
 \mu&=&\sqrt{M^2_{X/Y/Z}-(2{\mathbb{M}}_c)^2}\, ,
 \end{eqnarray}
 where the effective mass of the $c$-quark ${\mathbb{M}}_c=1.8\,\rm{GeV}$.
  The heavy tetraquark system could be described
by a double-well potential with two light quarks $q^{\prime}\bar{q}$ lying in the two wells respectively.
   In the heavy quark limit, the $c$ (and $b$) quark can be taken as a static well potential,
which binds the light quark $q$ to form a diquark in the color antitriplet channel. The heavy tetraquark state are characterized by the effective heavy quark masses ${\mathbb{M}}_Q$ (or constituent quark masses) and the virtuality $V=\sqrt{M^2_{X/Y/Z}-(2{\mathbb{M}}_Q)^2}$ (or bound energy not as robust). It is natural to take the energy  scale $\mu=V$.
  The energy scales are estimated to be $\mu=1.5\,\rm{GeV}$ for the $X(3872)$ and $Z_c(3900)$ \cite{WangHuangTao}, $\mu=3.0\,\rm{GeV}$ for the $Y(4660)$, and $\mu=2.5\,\rm{GeV}$ for the $Y(4360)$.
 The formula also works well for the scalar hidden charmed (and double charmed) tetraquark states, and we can use the formula to
   improve the predictions  \cite{cc-Wang-scalar}. Furthermore, we study the possible applications in the QCD sum rules for the molecular states \cite{Wang4140}.
 From Fig.1, we can see that the energy scales $\mu=2.5\,\rm{GeV}$ and $3.0\,\rm{GeV}$ lead to slightly different masses for the threshold parameters  $\sqrt{s_0}=4.7\,\rm{GeV}$ or larger than $4.7\,\rm{GeV}$. In this article, we set the energy scale $\mu=3.0\,\rm{GeV}$ to study the vector tetraquark states.

In Fig.2,  the contributions of the pole terms are plotted with
variations of the threshold parameters $\sqrt{s_0}$ and Borel parameters $T^2$ at the energy scale $\mu=3.0\,\rm{GeV}$.
From the figure, we can
see that the values $\sqrt{s_0}\leq 4.8 \, \rm{GeV}$ are too small to
satisfy the pole dominance condition and result in reasonable Borel windows.
In Fig.3,  the contributions of different terms in the
operator product expansion are plotted with variations of the Borel parameters  $T^2$ for the threshold parameters $\sqrt{s_0}=5.1\,\rm{GeV}$ and $5.0\,\rm{GeV}$ in the channels $c\bar{c}s\bar{s}$ and $c\bar{c}(u\bar{u}+d\bar{d})/\sqrt{2}$  respectively at the energy scale $\mu=3.0\,\rm{GeV}$. From the figure, we can see that the contributions of the vacuum condensates of dimensions-0, 5, 6 change quickly with variations of the Borel parameters
at the region $T^2< 3.2\,\rm{GeV}^2$, which does  not warrant
platforms for the masses. In this article,  the
value $T^2\geq 3.2\,\rm{GeV}^2$ is chosen tentatively, in that case  the convergent
behavior in the operator product  expansion is very good, as the perturbative terms make the main contributions.
The Borel parameters, continuum threshold parameters and the pole contributions are shown explicitly in Table 1. The two criteria (pole dominance and convergence of the operator product expansion) of the QCD sum rules are fully satisfied, so we expect to make reasonable predictions. While in the QCD sum rules for the light tetraquark states, the two criteria are difficult to satisfy \cite{Wang-NPA}.

\begin{table}
\begin{center}
\begin{tabular}{|c|c|c|c|c|c|c|c|}\hline\hline
                               & $T^2 (\rm{GeV}^2)$ & $\sqrt{s_0} (\rm{GeV})$ & pole         & $M_{Y/Z}(\rm{GeV})$       & $\lambda_{Y/Z}(10^{-2}\rm{GeV}^5)$ \\ \hline
 $c\bar{c}s\bar{s}$ ($1^{-+}$) & $3.4-3.8$          & $5.1\pm0.1$             & $(47-66)\%$  & $4.63^{+0.11}_{-0.08}$    & $6.82^{+0.99}_{-0.80}$       \\ \hline
 $c\bar{c}u\bar{d}$ ($1^{-+}$) & $3.2-3.6$          & $5.0\pm0.1$             & $(48-67)\%$  & $4.57^{+0.12}_{-0.08}$    & $6.26^{+1.05}_{-0.79}$    \\ \hline
 $c\bar{c}s\bar{s}$ ($1^{--}$) & $3.4-3.8$          & $5.1\pm0.1$             & $(44-63)\%$  & $4.70^{+0.14}_{-0.10}$    & $7.08^{+1.29}_{-0.93}$    \\ \hline
 $c\bar{c}u\bar{d}$ ($1^{--}$) & $3.2-3.6$          & $5.0\pm0.1$             & $(44-64)\%$  & $4.66^{+0.17}_{-0.10}$    & $6.60^{+1.54}_{-0.95}$    \\ \hline
 \hline
\end{tabular}
\end{center}
\caption{ The Borel parameters, continuum threshold parameters, pole contributions, masses and pole residues of the vector tetraquark states. }
\end{table}

\begin{figure}
\centering
\includegraphics[totalheight=6cm,width=7cm]{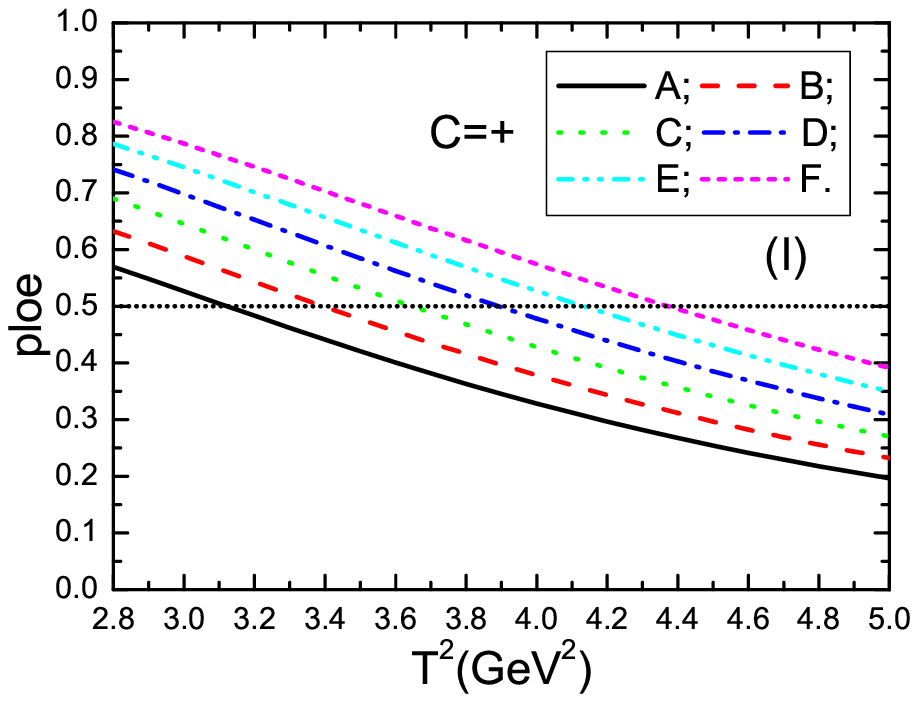}
\includegraphics[totalheight=6cm,width=7cm]{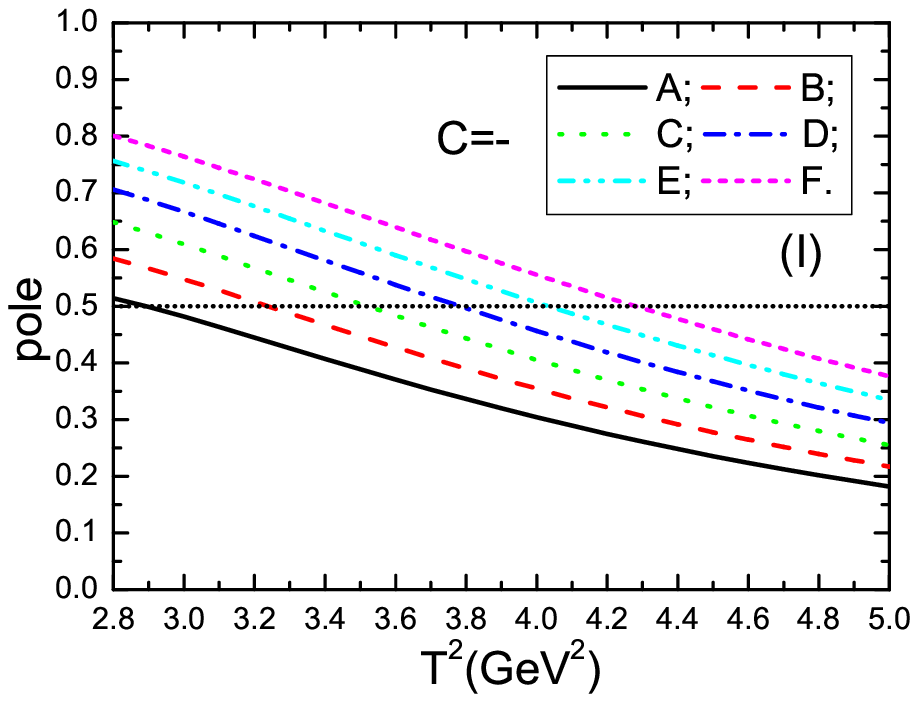}
\includegraphics[totalheight=6cm,width=7cm]{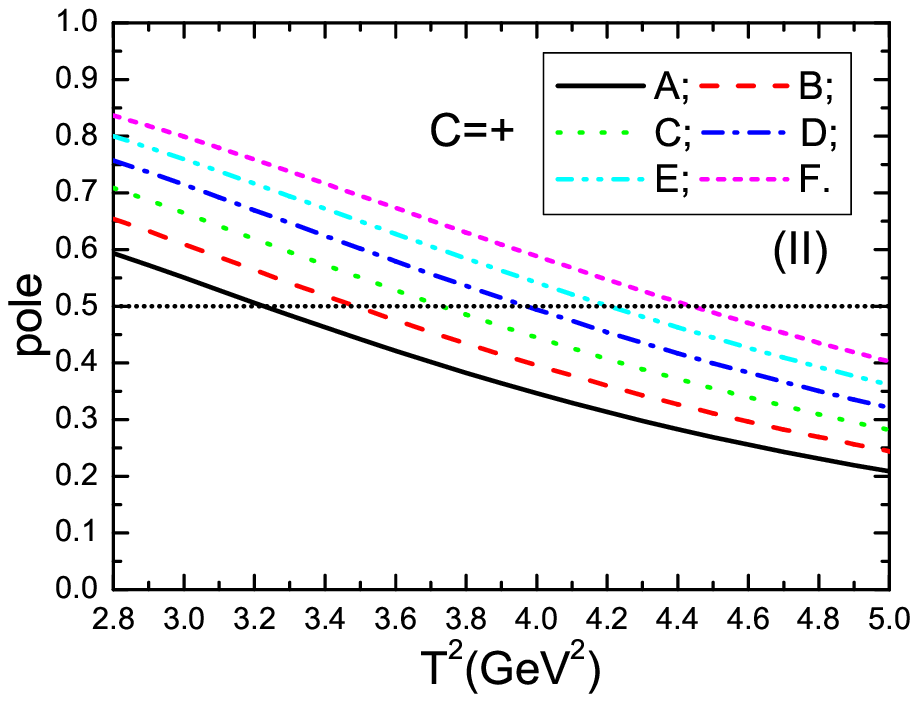}
\includegraphics[totalheight=6cm,width=7cm]{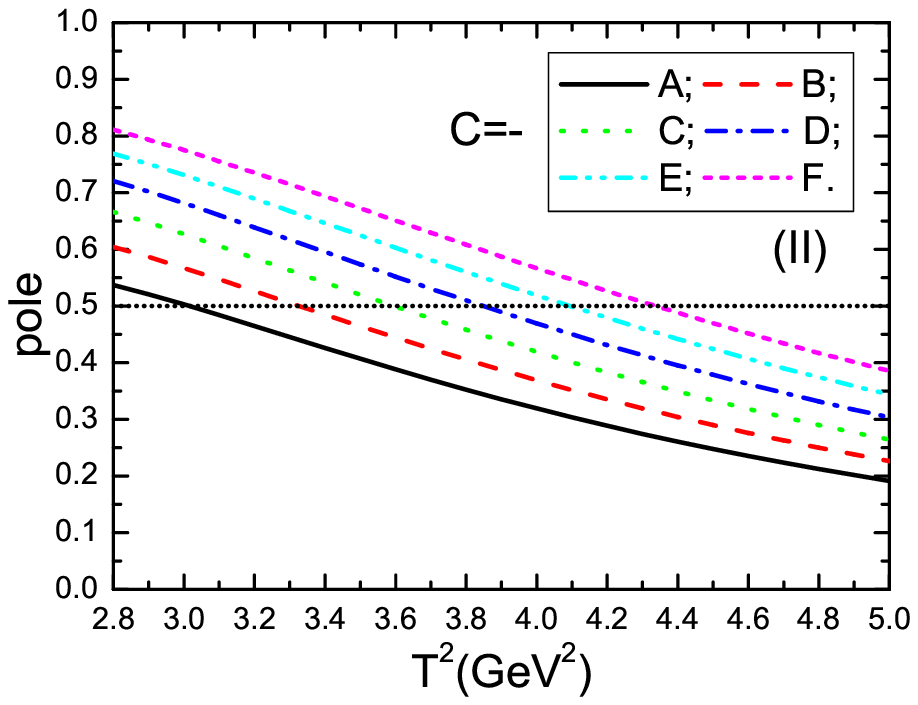}
  \caption{ The pole contributions  with variations of the  Borel parameters $T^2$ and threshold parameters $\sqrt{s_0}$, where the $A$, $B$, $C$, $D$, $E$ and $F$ denote the threshold parameters $\sqrt{s_0}=4.8$, 4.9, 5.0, 5.1, 5.2 and $5.3\,\rm{GeV}$, respectively; the (I) and (II) denote the $c\bar{c}s\bar{s}$ and $c\bar{c}(u\bar{u}+d\bar{d})/\sqrt{2}$ tetraquark states, respectively; the $C=\pm$ denote the charge conjugations; the horizontal lines denote the pole contributions of $50\%$. }
\end{figure}
\begin{figure}
\centering
\includegraphics[totalheight=6cm,width=7cm]{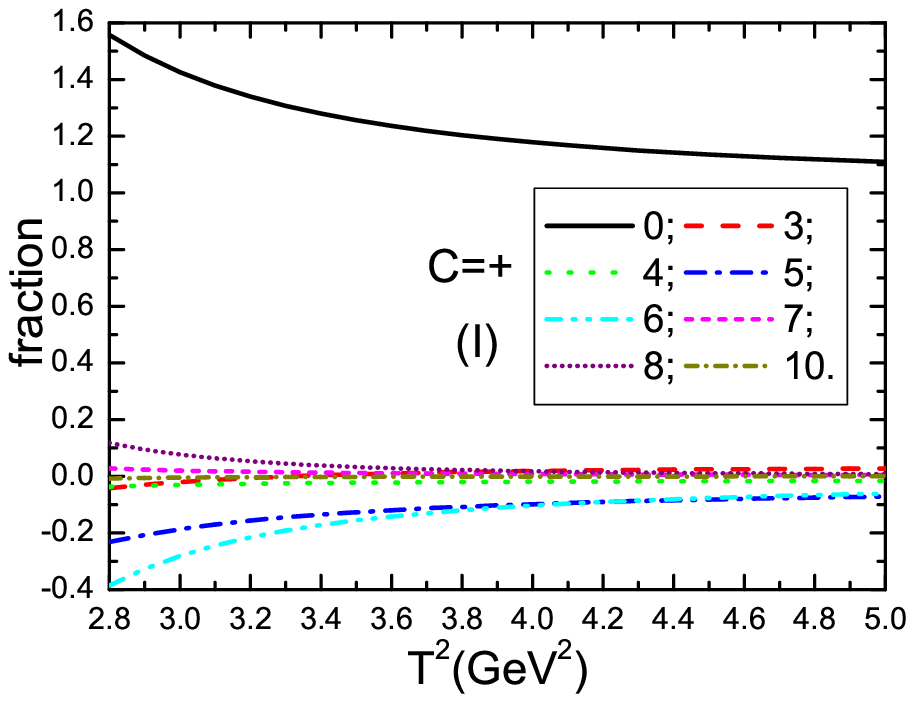}
\includegraphics[totalheight=6cm,width=7cm]{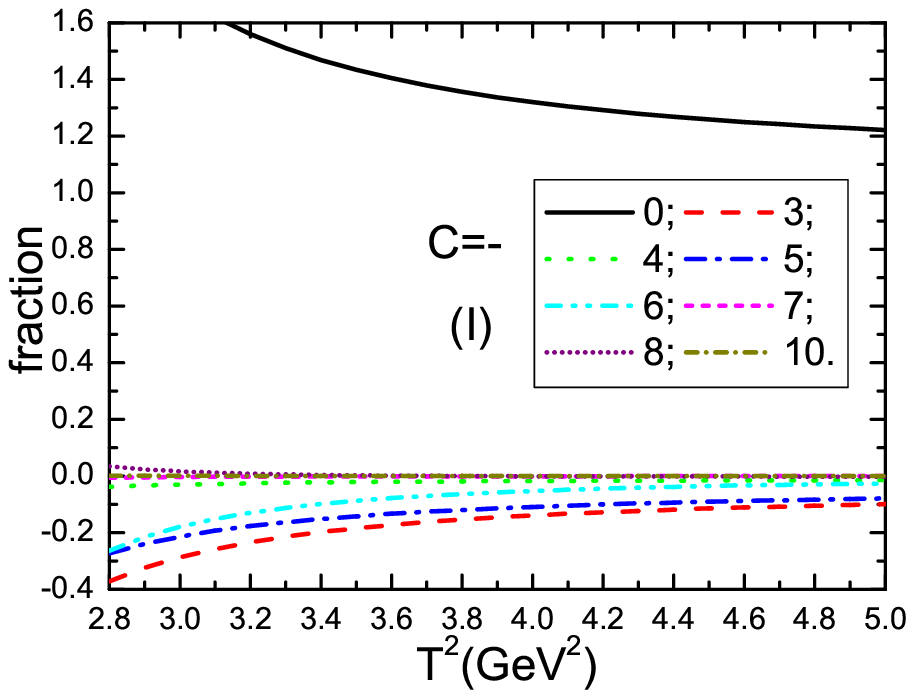}
\includegraphics[totalheight=6cm,width=7cm]{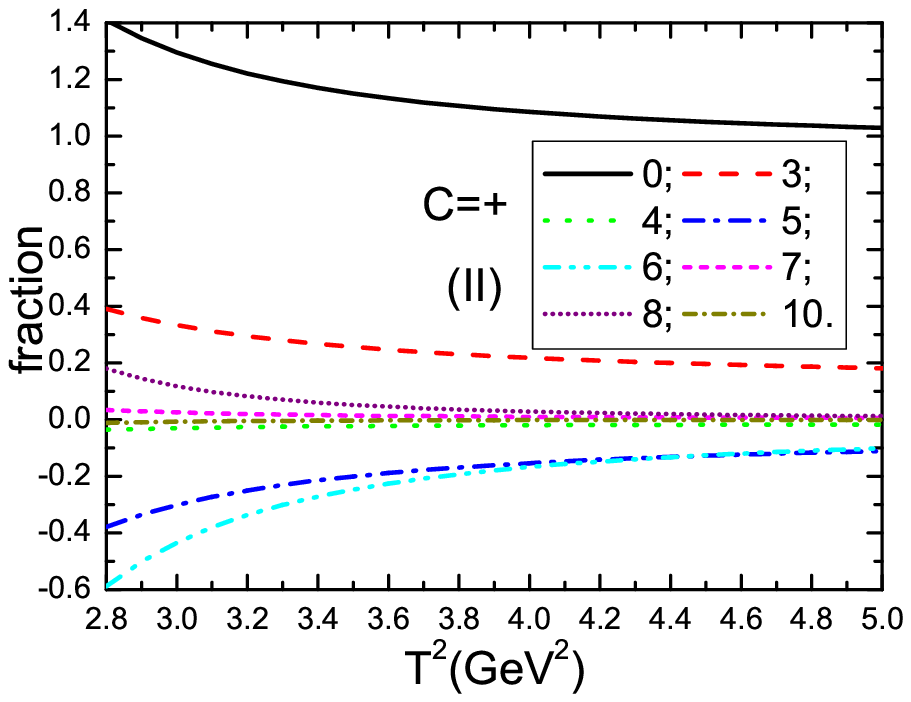}
\includegraphics[totalheight=6cm,width=7cm]{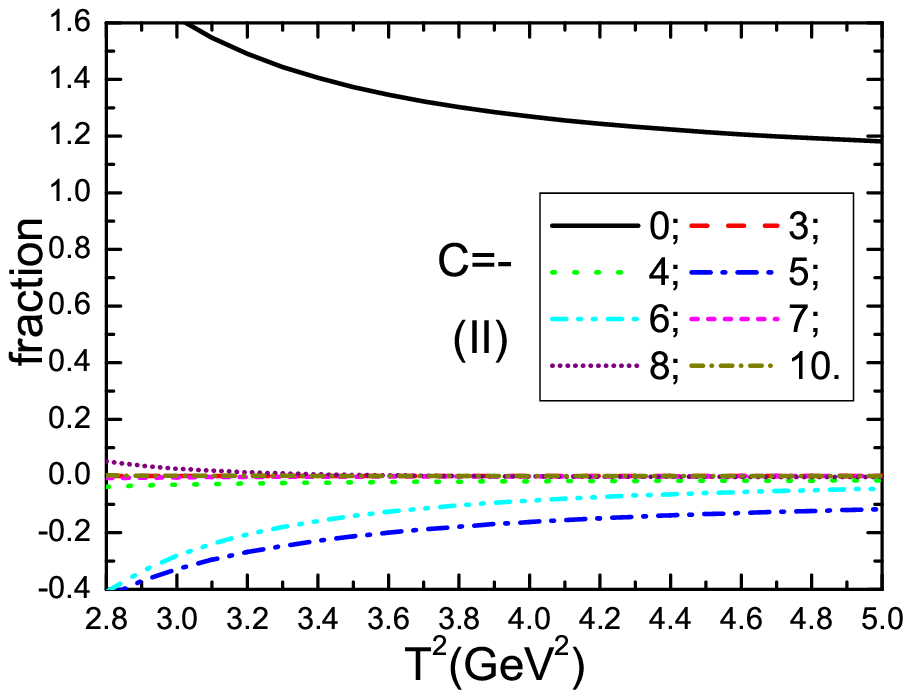}
  \caption{ The contributions of different terms in the operator product expansion  with variations of the  Borel parameters $T^2$, where the 0, 3, 4, 5, 6 ,7, 8 and 10 denotes the dimensions of the vacuum condensates; the (I) and (II) denote the $c\bar{c}s\bar{s}$ and $c\bar{c}(u\bar{u}+d\bar{d})/\sqrt{2}$ tetraquark states, respectively; the $C=\pm$ denote the charge conjugations. }
\end{figure}

Taking into account all uncertainties of the input parameters,
finally we obtain the values of the masses and pole residues of
 the   vector tetraquark states, which are  shown explicitly in Figs.4-5 and Table 1.
The prediction  $M_{c\bar{c}s\bar{s} (1^{--})}=4.70^{+0.14}_{-0.10}\,\rm{GeV}$ is consistent with the experimental data $M_{Y(4660)}=(4664\pm11\pm5)\,\rm{MeV}$ within uncertainties \cite{PDG}, and the prediction  $M_{c\bar{c}u\bar{d} (1^{--})}$ or $M_{c\bar{c}(u\bar{u}+d\bar{d})/\sqrt{2} (1^{--})}$ $=4.66^{+0.17}_{-0.10}\,\rm{GeV}$ is much larger than upper bound of the experimental data $M_{Y(4360)}=(4361\pm 9\pm9)\,\rm{ MeV}$ \cite{PDG}. The present predictions favor assigning the $Y(4660)$   as the $J^{PC}=1^{--}$   diquark-antidiquark type tetraquark state, the masses $M_{c\bar{c}s\bar{s}}$ and $M_{c\bar{c}(u\bar{u}+d\bar{d})/\sqrt{2} }$ are both consistent with the experimental data $M_{Y(4660)}$ within uncertainties. By precisely measuring the $\pi^+\pi^-$ mass spectrum in the final state $\pi^+\pi^-\psi^{\prime}$, we can distinguish the $f_0(600)$ and $f_0(980)$, therefore disentangle   the quark constituents of the $Y(4660)$. On the other hand, we can also take the $Y(4360)$ as the $c\bar{c}$ and $c\bar{c}(u\bar{u}+d\bar{d})/\sqrt{2}$ mixed state, as the $c\bar{c}$ component can reduce the mass so as  to reproduce the experimental value at about $4.4\,\rm{GeV}$.

From Table 1, we can also see that there is energy gap about $(70-90)\,\rm{MeV}$ between the central values of the  $C=+$ and $C=-$ vector tetraquark states, which can be confronted with the experimental data in the future.
In Ref.\cite{WangHuangTao}, we observe that there is a small energy gap less than $40\,\rm{MeV}$ between the central values of the $C=+$ and $C=-$ axial-vector tetraquark states, which is consistent with the value $10\,\rm{MeV}$ from the constituent diquark model \cite{Maiani-3872}.

\begin{figure}
\centering
\includegraphics[totalheight=6cm,width=7cm]{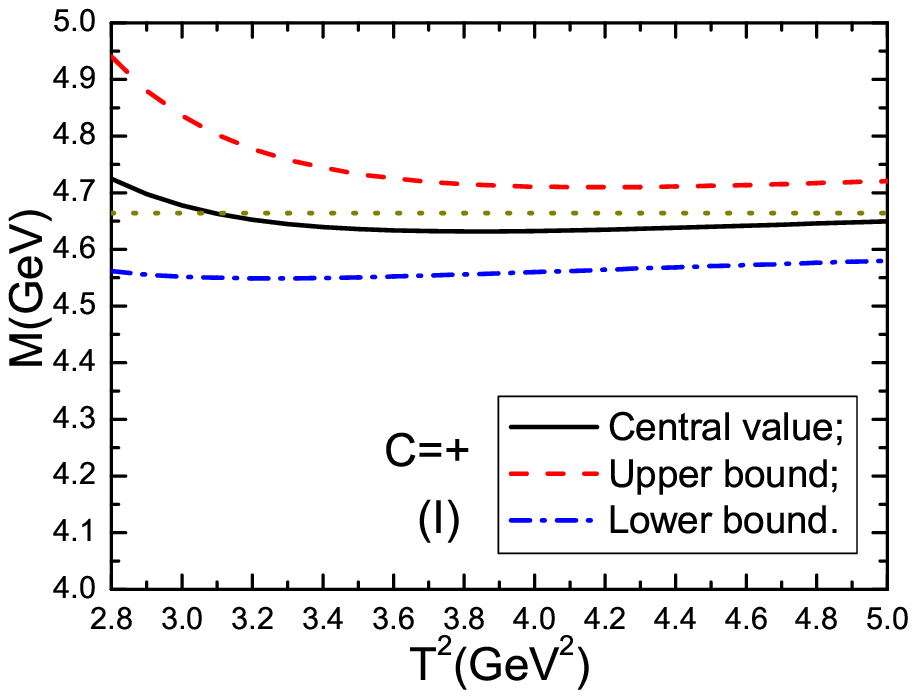}
\includegraphics[totalheight=6cm,width=7cm]{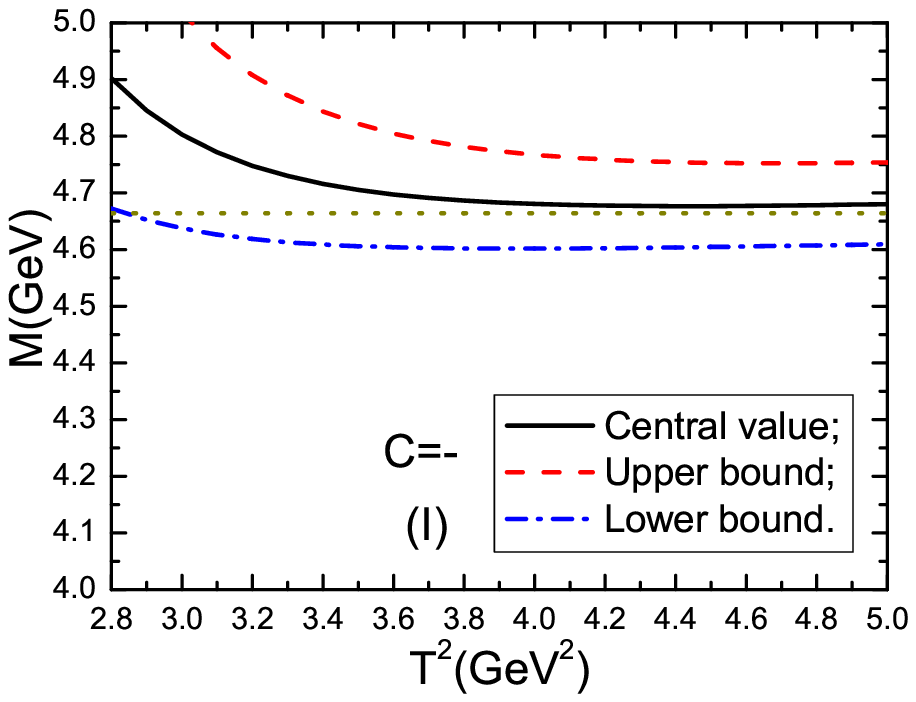}
\includegraphics[totalheight=6cm,width=7cm]{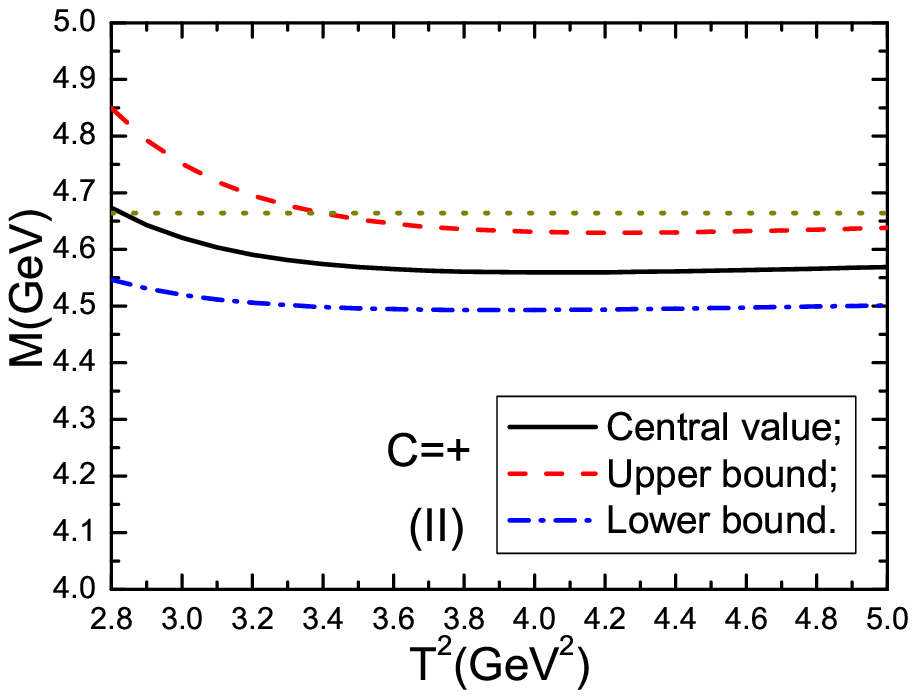}
\includegraphics[totalheight=6cm,width=7cm]{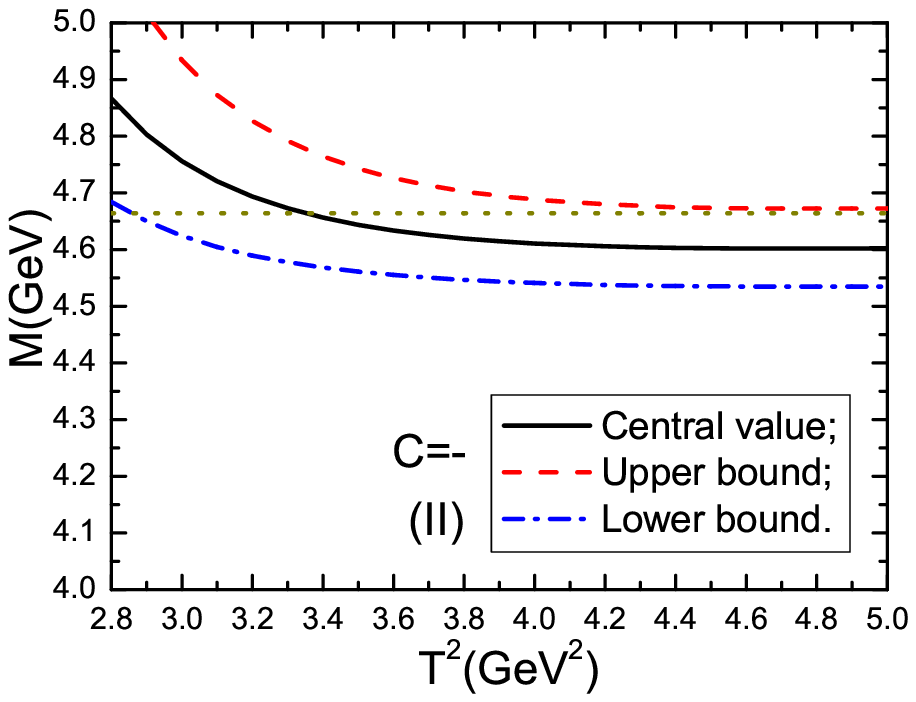}
  \caption{ The masses  with variations of the  Borel parameters $T^2$, where the horizontal lines denote the experimental value of the mass of  the $Y(4660)$;
  the (I) and (II) denote the $c\bar{c}s\bar{s}$ and $c\bar{c}(u\bar{u}+d\bar{d})/\sqrt{2}$ tetraquark states, respectively; the $C=\pm$ denote the charge conjugations. }
\end{figure}
\begin{figure}
\centering
\includegraphics[totalheight=6cm,width=7cm]{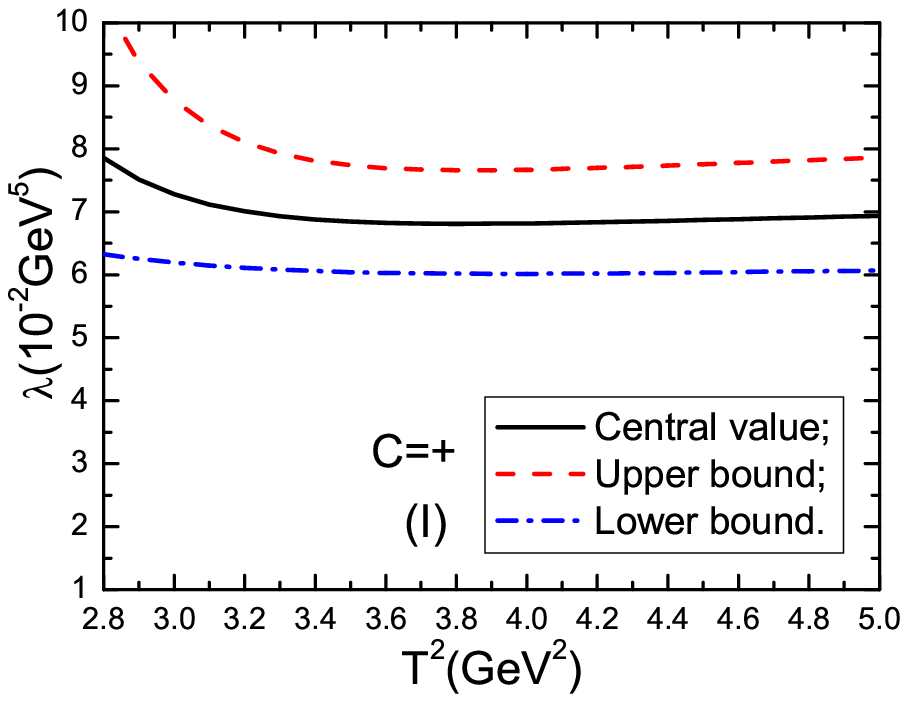}
\includegraphics[totalheight=6cm,width=7cm]{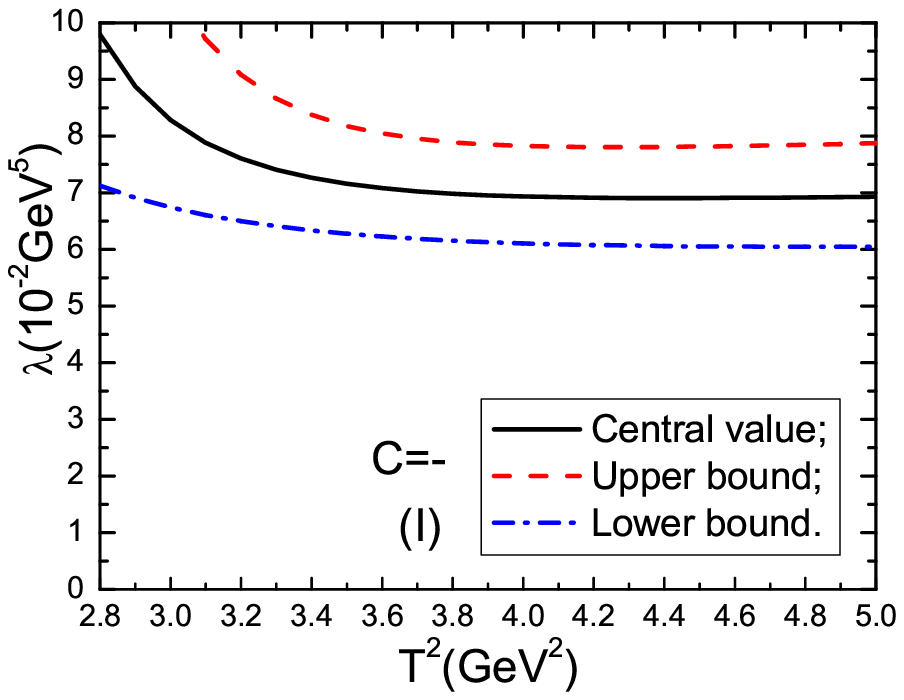}
\includegraphics[totalheight=6cm,width=7cm]{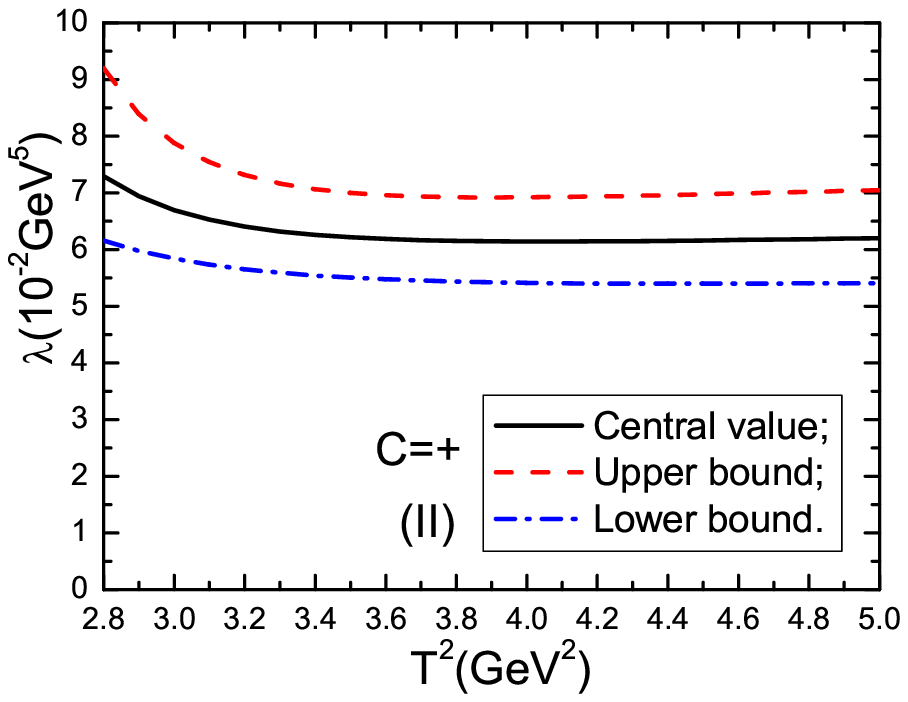}
\includegraphics[totalheight=6cm,width=7cm]{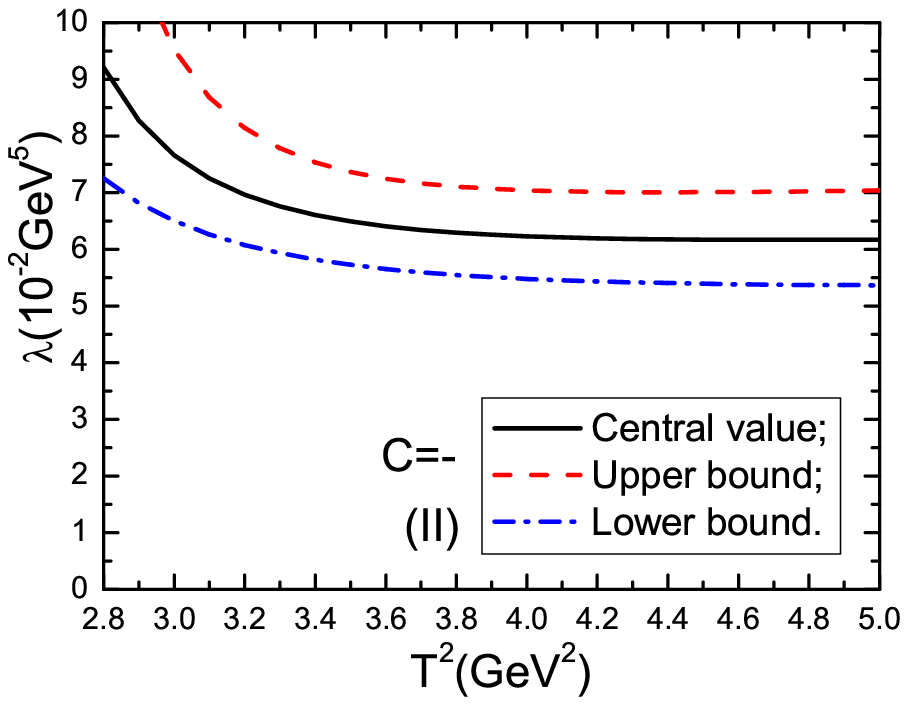}
  \caption{ The pole residues  with variations of the  Borel parameters $T^2$, where the (I) and (II) denote the $c\bar{c}s\bar{s}$ and $c\bar{c}(u\bar{u}+d\bar{d})/\sqrt{2}$ tetraquark states, respectively; the $C=\pm$ denote the charge conjugations. }
\end{figure}

In this article, we construct the $C-C\gamma_\mu$ type   diquark-antidiquark  currents to interpolate the vector tetraquark states. The scalar and axial-vector heavy-light diquark states have almost  degenerate masses from the QCD sum rules \cite{WangDiquark}, the $C\gamma_\mu-C$ and $C\gamma_\mu\gamma_5-C\gamma_5$ type   tetraquark states  have degenerate (or slightly different) masses \cite{WangJPG}. On the other hand, we can also construct
the $C\gamma_\alpha-\partial_\mu-C\gamma^\alpha$ and $C\gamma_5-\partial_\mu-C\gamma_5$ type diquark-antidiquark  currents to interpolate the vector tetraquark states, the $C\gamma_\alpha-C\gamma^\alpha$ and $C\gamma_5-C\gamma_5$ type diquark-antidiquark currents couple    to the scalar tetraquark states with the masses about $3.85\,\rm{GeV}$ \cite{Wangzg1312}. If the contribution of an additional P-wave  to the mass is about $0.5\,\rm{GeV}$, the masses of the vector tetraquark states couple to the  $C\gamma_\alpha-\partial_\mu-C\gamma^\alpha$ and $C\gamma_5-\partial_\mu-C\gamma_5$ type interpolating  currents are about $4.35\,\rm{GeV}$, which happens to be the value of the mass  $M_{Y(4360)}$. In Refs.\cite{Tetraquark-Ebert,ZhangHuang-JHEP}, Zhang and Huang take the $C\gamma_5-\partial_\mu-C\gamma_5$ type diquark-antidiquark  currents to study the $Y(4360)$ and $Y(4660)$ with the symbolic quark structures  $c\bar{c}(u\bar{u}+d\bar{d})/\sqrt{2}$ and $c\bar{c}s\bar{s}$, respectively,  and obtain the values $M_{Y(4360)}=(4.32\pm0.20)\,\rm{GeV}$ and $M_{Y(4660)}=(4.69\pm0.36)\,\rm{GeV}$, which are consistent with the rough estimation $M_{Y(4360)}=4.35\,\rm{GeV}$. The present predictions
 $M_{c\bar{c}u\bar{d}(1^{-+})}=4.57^{+0.12}_{-0.08}\,\rm{GeV}$ and     $M_{c\bar{c}u\bar{d} (1^{--})}=4.66^{+0.17}_{-0.10}\,\rm{GeV}$
disfavor assigning the $Z_c(4025)$ and $Z_c(4020)$ as the $J^{PC}=1^{--}$ tetraquark states, and favor assigning the $Y(4360)$ as the $C\gamma_\alpha-\partial_\mu-C\gamma^\alpha$ and $C\gamma_5-\partial_\mu-C\gamma_5$ type $J^{PC}=1^{--}$ tetraquark states.

Now we perform Fierz re-arrangement  to the vector currents $J_{1^{--},\bar{d}u}^{\mu}$, $J_{1^{-+},\bar{d}u}^{\mu}$, $J_{1^{--},\bar{s}s}^{\mu}$,
  $J_{1^{-+},\bar{s}s}^{\mu}$ both in the color and Dirac-spinor  spaces,  and obtain the following results,
 \begin{eqnarray}
J_{1^{--},\bar{d}u}^{\mu}&=&\frac{\epsilon^{ijk}\epsilon^{imn}}{\sqrt{2}}\left\{u^jC c^k \bar{d}^m\gamma^\mu C \bar{c}^n-u^jC\gamma^\mu c^k\bar{d}^m C \bar{c}^n \right\} \, , \nonumber\\
 &=&\frac{1}{2\sqrt{2}}\left\{\,\bar{c} \gamma^\mu c\,\bar{d} u-\bar{c} c\,\bar{d}\gamma^\mu u+i\bar{c}\gamma^\mu\gamma_5 u\,\bar{d}i\gamma_5 c-i\bar{c} i\gamma_5 u\,\bar{d}\gamma^\mu \gamma_5c\right. \nonumber\\
&&\left. - i\bar{c}\gamma_\nu\gamma_5c\, \bar{d}\sigma^{\mu\nu}\gamma_5u+i\bar{c}\sigma^{\mu\nu}\gamma_5c\, \bar{d}\gamma_\nu\gamma_5u
- i\bar{d}\gamma_\nu c\, \bar{c}\sigma^{\mu\nu}u+i \bar{d}\sigma^{\mu\nu}c \,\bar{c}\gamma_\nu u  \,\right\} \, , \\
J_{1^{-+},\bar{d}u}^{\mu}&=&\frac{\epsilon^{ijk}\epsilon^{imn}}{\sqrt{2}}\left\{u^jC c^k \bar{d}^m\gamma^\mu C \bar{c}^n+u^jC\gamma^\mu c^k\bar{d}^m C \bar{c}^n \right\} \, , \nonumber\\
 &=&\frac{1}{2\sqrt{2}}\left\{\,i\bar{c}i\gamma_5 c\,\bar{d}\gamma^\mu\gamma_5 u-i\bar{c} \gamma^\mu\gamma_5 c\,\bar{d}i\gamma_5 u-\bar{c}\gamma^\mu u\,\bar{d} c+\bar{c} u\,\bar{d}\gamma^\mu c\right. \nonumber\\
&&\left.+i\bar{c}\sigma^{\mu\nu} c\, \bar{d}\gamma_\nu u - i\bar{c}\gamma_\nu c\, \bar{d}\sigma^{\mu\nu} u
- i\bar{d}\gamma_\nu\gamma_5 c\, \bar{c}\sigma^{\mu\nu}\gamma_5u+i \bar{d}\sigma^{\mu\nu}\gamma_5 c \,\bar{c}\gamma_\nu\gamma_5 u  \,\right\} \, ,
\end{eqnarray}
\begin{eqnarray}
J_{1^{--},\bar{s}s}^{\mu}&=&\frac{\epsilon^{ijk}\epsilon^{imn}}{\sqrt{2}}\left\{s^jC c^k \bar{s}^m\gamma^\mu C \bar{c}^n-s^jC\gamma^\mu c^k\bar{s}^m C \bar{c}^n \right\} \, , \nonumber\\
 &=&\frac{1}{2\sqrt{2}}\left\{\,\bar{c} \gamma^\mu c\,\bar{s} s-\bar{c} c\,\bar{s}\gamma^\mu s+i\bar{c}\gamma^\mu\gamma_5 s\,\bar{s}i\gamma_5 c-i\bar{c} i\gamma_5 s\,\bar{s}\gamma^\mu \gamma_5c\right. \nonumber\\
&&\left. - i\bar{c}\gamma_\nu\gamma_5c\, \bar{s}\sigma^{\mu\nu}\gamma_5s+i\bar{c}\sigma^{\mu\nu}\gamma_5c\, \bar{s}\gamma_\nu\gamma_5s
- i\bar{s}\gamma_\nu c\, \bar{c}\sigma^{\mu\nu}s+i \bar{s}\sigma^{\mu\nu}c \,\bar{c}\gamma_\nu s  \,\right\} \, , \\
J_{1^{-+},\bar{s}s}^{\mu}&=&\frac{\epsilon^{ijk}\epsilon^{imn}}{\sqrt{2}}\left\{s^jC c^k \bar{s}^m\gamma^\mu C \bar{c}^n+s^jC\gamma^\mu c^k\bar{s}^m C \bar{c}^n \right\} \, , \nonumber\\
 &=&\frac{1}{2\sqrt{2}}\left\{\,i\bar{c}i\gamma_5 c\,\bar{s}\gamma^\mu\gamma_5 s-i\bar{c} \gamma^\mu\gamma_5 c\,\bar{s}i\gamma_5 s-\bar{c}\gamma^\mu s\,\bar{s} c+\bar{c} s\,\bar{s}\gamma^\mu c\right. \nonumber\\
&&\left.+i\bar{c}\sigma^{\mu\nu} c\, \bar{s}\gamma_\nu s - i\bar{c}\gamma_\nu c\, \bar{s}\sigma^{\mu\nu} s
- i\bar{s}\gamma_\nu\gamma_5 c\, \bar{c}\sigma^{\mu\nu}\gamma_5s+i \bar{s}\sigma^{\mu\nu}\gamma_5 c \,\bar{c}\gamma_\nu\gamma_5 s  \,\right\} \, ,
\end{eqnarray}
where  the subscripts $1^{-\pm}$ and $\bar{d}u$ ($\bar{s}s$) are added to show the $J^{PC}$ and light quark constituents, respectively.
Then we obtain the OZI super-allowed decays by taking into account the couplings to the meson-meson pairs,
\begin{eqnarray}
Y(cu\bar{c}\bar{d})(1^{--})&\to& J/\psi a_0(980)\, , \,\chi_{c0} \rho\, , \, \chi_{c1} \rho\, , \,J/\psi \pi\, , \,J/\psi a_1(1260)\, , \, DD_{1}(2420)\, ,\, D^*D_{1}(2420)\, ,\nonumber\\
Y(cu\bar{c}\bar{d})(1^{-+})&\to& \eta_c \pi\, , \,\eta_c a_1(1260)\, , \,\chi_{c1} \pi\, , \,h_c \rho\, , \,J/\psi b_1(1235)\, , \, D^*D_{0}(2400)\, ,\, D^*D_{1}(2420)\, ,\nonumber\\
\end{eqnarray}
\begin{eqnarray}
Y(4660)(1^{--})&\to& J/\psi f_0(980)\, , \,J/\psi f_0(600)\, , \,J/\psi (\pi\pi)_S\, , \, \psi^{\prime} f_0(980)\, , \,\psi^{\prime} f_0(600)\, , \,\psi^{\prime} (\pi\pi)_S\, , \,\nonumber\\
&&\chi_{c0} \phi(1020)\, , \, \chi_{c0} (KK)_{S}\, , \,\chi_{c1} \phi(1020)\, , \,\chi_{c1} (KK)_{S}\, , \,J/\psi h_1(1380)/f_1(1510)\, , \,\nonumber\\
&&J/\psi\eta\, , \, D_sD_{s1}(2460)/D_{s1}(2536)\, ,\, D_s^*D_{s1}(2460)/D_{s1}(2536)\, ,\nonumber\\
Y(cs\bar{c}\bar{s})(1^{-+})&\to& \eta_c h_1(1380)/f_1(1510)\, , \,\eta_c \eta\, , \,\chi_{c1} \eta\, , \,h_c \phi(1020)\, , \,J/\psi h_1(1380)/f_1(1510)\, , \, \nonumber\\
&&D_s^*D_{s0}(2317)\, ,\, D_s^*D_{s1}(2460)/D_{s1}(2536)\, ,
\end{eqnarray}
where the $(\pi\pi)_S$ and $(KK)_S$ denote the S-wave $\pi\pi$ and $KK$ pairs, respectively.

The mass spectrum of the light scalar mesons  is well understood in terms of diquark-antidiquark bound
states,  while the strong decays into two pseudoscalar mesons  based on the quark rearrangement mechanism cannot lead to a satisfactory description of the experimental data. In Ref.\cite{Hooft0801}, 't Hooft et al introduce the instanton induced effective six-fermion Lagrangian, and illustrate that such Lagrangian
leads to the tetraquark-$\bar{q}q$ mixing, therefore provides an additional amplitude which brings the strong decays of the light scalar mesons in good agreements  with
the experimental data. In the present work, we discuss the OZI super-allowed strong decays of the tetraquark states based on the quark rearrangement mechanism or fall-apart mechanism,
as there is no instanton induced effective six-fermion Lagrangian in the hidden-charm systems to describe the tetraquark-$\bar{q}q$ mixing beyond the usual QCD interactions. The present predictions can be confronted with the experimental data in the futures  the BESIII, LHCb and Belle-II.

\section{Conclusion}
In this article, we study the $Z_c(4020)$, $Z_c(4025)$, $Y(4360)$ and $Y(4660)$ as the diquark-antidiquark type vector tetraquark states in details with the QCD sum rules. We distinguish the charge conjugations of the interpolating currents,  calculate the contributions of the vacuum condensates up to dimension-10  and discard the perturbative corrections in the operator product expansion, and  take into account  the higher dimensional vacuum condensates consistently, as they play an important role in determining the Borel windows. Then we suggest a  formula $\mu=\sqrt{M^2_{X/Y/Z}-(2{\mathbb{M}}_c)^2}$ to estimate the energy scales of the QCD spectral densities
of the hidden charmed tetraquark states, and study  the masses and pole residues of the $J^{PC}=1^{-\pm}$  tetraquark states in details. The
 formula $\mu=\sqrt{M^2_{X/Y/Z}-(2{\mathbb{M}}_c)^2}$ works well. The   masses
of the $c\bar{c}u\bar{d}$ ($1^{-\pm}$)  tetraquark states disfavor assigning  the $Z_c(4020)$, $Z_c(4025)$ and $Y(4360)$
as the  $C-C\gamma_\mu$ type vector tetraquark states, and favor assigning  the $Z_c(4020)$, $Z_c(4025)$ as the
diquark-antidiquark  type $1^{+-}$ tetraquark states. While the masses of the $c\bar{c}s\bar{s}$ and $c\bar{c}(u\bar{u}+d\bar{d})/\sqrt{2}$ tetraquark states favor
 assigning the $Y(4660)$ as the $C-C\gamma_\mu$ type $1^{--}$ tetraquark state, more experimental data are still needed to distinguish the quark constituents.
There are no candidates for   the $C=+$ vector tetraquark states, the predictions can be confronted with the experimental data in the futures at the BESIII, LHCb and Belle-II. The  pole residues can be taken as   basic input parameters to study relevant processes of the vector tetraquark states with the three-point QCD sum rules.

\section*{Appendix}
The spectral densities $\rho_i(s)$ with $i=0$, 3, 4, 5, 6, 7, 8, 10  at the level of the quark-gluon degrees of
freedom,

\begin{eqnarray}
\rho_{0}(s)&=&\frac{1}{3072\pi^6}\int dydz \, yz(1-y-z)^3\left(s-\overline{m}_c^2\right)^2\left(35s^2-26s\overline{m}_c^2+3\overline{m}_c^2 \right)  \nonumber\\
&&-t\frac{m_c^2}{1536\pi^6} \int dydz \, (1-y-z)^3\left(s-\overline{m}_c^2\right)^3  \nonumber\\
&&+(1+t)\frac{m_s m_c}{512\pi^6} \int dydz \, (y+z) (1-y-z)^2\left(s-\overline{m}_c^2\right)^3 \, ,
\end{eqnarray}

\begin{eqnarray}
\rho_{3}(s)&=&-(1+t)\frac{m_c\langle \bar{s}s\rangle}{64\pi^4}\int dydz \, (y+z)(1-y-z)\left(s-\overline{m}_c^2\right)^2  \nonumber \\
&&+\frac{m_s\langle \bar{s}s\rangle}{32\pi^4}\int dydz \, yz\,(1-y-z)\left(15s^2-16s\overline{m}_c^2+3\overline{m}_c^4\right)+\frac{m_s m_c^2\langle \bar{s}s\rangle}{8\pi^4}\int dydz \left(s-\overline{m}_c^2\right)   \nonumber\\
&&+t\frac{m_s  \langle \bar{s}s\rangle}{32\pi^4}\int dydz \, yz \left(s-\overline{m}_c^2\right)^2 -t\frac{m_s m_c^2\langle \bar{s}s\rangle}{32\pi^4}\int dydz \,(1-y-z)\left(s-\overline{m}_c^2\right)  \, ,
\end{eqnarray}

\begin{eqnarray}
\rho_{4}(s)&=&-\frac{m_c^2}{2304\pi^4} \langle\frac{\alpha_s GG}{\pi}\rangle\int dydz \left( \frac{z}{y^2}+\frac{y}{z^2}\right)(1-y-z)^3 \left\{ 8s-3\overline{m}_c^2+\overline{m}_c^4\delta\left(s-\overline{m}_c^2\right)\right\} \nonumber\\
&&+t\frac{m_c^4}{4608\pi^4}\langle\frac{\alpha_s GG}{\pi}\rangle\int dydz \left(\frac{1}{y^3}+\frac{1}{z^3} \right) (1-y-z)^3 \nonumber\\
&&-t\frac{m_c^2}{1536\pi^4}\langle\frac{\alpha_s GG}{\pi}\rangle\int dydz \left(\frac{1}{y^2}+\frac{1}{z^2} \right) (1-y-z)^3 \left(s-\overline{m}_c^2\right) \nonumber\\
&&+\frac{1}{1536\pi^4}\langle\frac{\alpha_s GG}{\pi}\rangle\int dydz \, (y+z)   (1-y-z)^2 \,s\,\left(5s-4\overline{m}_c^2\right) \nonumber\\
&&-t\frac{m_c^2}{1024\pi^4}\langle\frac{\alpha_s GG}{\pi}\rangle\int dydz \left( \frac{1}{y}+\frac{1}{z}\right)(1-y-z)^2\left(s-\overline{m}_c^2 \right)  \nonumber \\
&&+t\frac{m_c^2}{2304\pi^4}\langle\frac{\alpha_s GG}{\pi}\rangle\int dydz \left\{ \frac{(1-y-z)^2}{yz}-\frac{(1-y-z)^3}{6yz}\right\}\left(s-\overline{m}_c^2 \right) \, ,
\end{eqnarray}

\begin{eqnarray}
\rho_{5}(s)&=&(1+t)\frac{m_c\langle \bar{s}g_s\sigma Gs\rangle}{128\pi^4}\int dydz  \, (y+z) \left(s-\overline{m}_c^2 \right) \nonumber\\
&&+\frac{m_c\langle \bar{s}g_s\sigma Gs\rangle}{128\pi^4}\int dydz   \left(\frac{y}{z}+\frac{z}{y} \right)(1-y-z) \left(2s-\overline{m}_c^2 \right)  \nonumber\\
&&-t\frac{m_c\langle \bar{s}g_s\sigma Gs\rangle}{128\pi^4}\int dydz \,  (1-y-z) \left(s-\overline{m}_c^2 \right)  \nonumber\\
&&-t\frac{m_c\langle \bar{s}g_s\sigma Gs\rangle}{1152\pi^4}\int dydz   \left(\frac{y}{z}+\frac{z}{y} \right)(1-y-z) \left(5s-3\overline{m}_c^2 \right)   \nonumber \\
&&-\frac{m_s\langle \bar{s}g_s\sigma Gs\rangle}{96\pi^4}\int dydz  \, yz \, \left\{8s-3\overline{m}_c^2 +\overline{m}_c^4 \delta\left(s-\overline{m}_c^2 \right) \right\}   \nonumber \\
&&-\frac{m_s m_c^2\langle \bar{s}g_s\sigma Gs\rangle}{32\pi^4}\int_{y_i}^{y_f}  dy  +t\frac{m_s m_c^2\langle \bar{s}g_s\sigma Gs\rangle}{192\pi^4}\int dydz \nonumber \\
&&-t\frac{m_s  \langle \bar{s}g_s\sigma Gs\rangle}{64\pi^4}\int_{y_i}^{y_f} dy \, y(1-y) \left(s-\widetilde{m}_c^2 \right)  \nonumber \\
&&+\frac{m_s m_c^2\langle \bar{s}g_s\sigma Gs\rangle}{128\pi^4}\int dydz \left( \frac{1}{y}+\frac{1}{z}\right)   +t\frac{m_s\langle \bar{s}g_s\sigma Gs\rangle}{256\pi^4}\int dydz \, (y+z)\left(s-\overline{m}_c^2 \right) \, ,
\end{eqnarray}

\begin{eqnarray}
\rho_{6}(s)&=&-\frac{m_c^2\langle\bar{s}s\rangle^2}{12\pi^2}\int_{y_i}^{y_f}dy -t\frac{ \langle\bar{s}s\rangle^2}{24\pi^2}\int_{y_i}^{y_f}dy\,y(1-y)\left( s-\widetilde{m}_c^2\right)\nonumber\\
&&+\frac{g_s^2\langle\bar{s}s\rangle^2}{648\pi^4}\int dydz\, yz \left\{8s-3\overline{m}_c^2 +\overline{m}_c^4\delta\left(s-\overline{m}_c^2 \right)\right\}-t\frac{g_s^2m_c^2\langle\bar{s}s\rangle^2}{1296\pi^4}\int dydz \nonumber\\
&&-\frac{g_s^2\langle\bar{s}s\rangle^2}{2592\pi^4}\int dydz\,(1-y-z)\left\{ \left(\frac{z}{y}+\frac{y}{z} \right)3\left(7s-4\overline{m}_c^2 \right)+\left(\frac{z}{y^2}+\frac{y}{z^2} \right)\right.\nonumber\\
&&\left.m_c^2\left[ 7+5\overline{m}_c^2\delta\left(s-\overline{m}_c^2 \right)\right]-(y+z)\left(4s-3\overline{m}_c^2 \right)-\left(\frac{1}{y}+\frac{1}{z} \right)t\frac{3m_c^2}{2}\right\} \nonumber\\
&&-\frac{g_s^2\langle\bar{s}s\rangle^2}{3888\pi^4}\int dydz\left\{  \left(\frac{z}{y}+\frac{y}{z} \right)3\left(2s-\overline{m}_c^2 \right)+\left(\frac{z}{y^2}+\frac{y}{z^2} \right)m_c^2\left[ 1+\overline{m}_c^2\delta\left(s-\overline{m}_c^2\right)\right]\right. \nonumber\\
&&\left.+(y+z)2\left[8s-3\overline{m}_c^2 +\overline{m}_c^4\delta\left(s-\overline{m}_c^2\right)\right]-\left(\frac{1}{y}+\frac{1}{z}\right)t\frac{3m_c^2}{2}\right\}\, ,
\end{eqnarray}

\begin{eqnarray}
\rho_7(s)&=&(1+t)\frac{m_c^3\langle\bar{s}s\rangle}{576\pi^2}\langle\frac{\alpha_sGG}{\pi}\rangle\int dydz \left(\frac{y}{z^3}+\frac{z}{y^3}+\frac{1}{y^2}+\frac{1}{z^2}\right)(1-y-z) \delta\left(s-\overline{m}_c^2\right)\nonumber\\
&&-(1+t)\frac{m_c\langle\bar{s}s\rangle}{192\pi^2}\langle\frac{\alpha_sGG}{\pi}\rangle\int dydz \left(\frac{y}{z^2}+\frac{z}{y^2}\right)(1-y-z)  \nonumber\\
&&-\frac{m_c\langle\bar{s}s\rangle}{64\pi^2}\langle\frac{\alpha_sGG}{\pi}\rangle\int dydz\left\{1+\frac{4\overline{m}_c^2}{9}\delta\left(s-\overline{m}_c^2\right) \right\} \nonumber\\
&&-t\frac{m_c\langle\bar{s}s\rangle}{384\pi^2}\langle\frac{\alpha_sGG}{\pi}\rangle\int dydz\left(\frac{z}{y}+\frac{y}{z}\right)
-(1+t)\frac{m_c\langle\bar{s}s\rangle}{1152\pi^2}\langle\frac{\alpha_sGG}{\pi}\rangle\int dydz \, ,
\end{eqnarray}

\begin{eqnarray}
\rho_8(s)&=&\frac{m_c^2\langle\bar{s}s\rangle\langle\bar{s}g_s\sigma Gs\rangle}{24\pi^2}\int_0^1 dy \left(1+\frac{\widetilde{m}_c^2}{T^2} \right)\delta\left(s-\widetilde{m}_c^2\right)\nonumber\\
&&+t\frac{ \langle\bar{s}s\rangle\langle\bar{s}g_s\sigma Gs\rangle}{16\pi^2}\int_{y_i}^{y_f} dy\, y(1-y)\left\{1+\frac{\widetilde{m}_c^2}{3} \delta\left(s-\widetilde{m}_c^2\right)\right\}\nonumber\\
&&-\frac{m_c^2\langle\bar{s}s\rangle\langle\bar{s}g_s\sigma Gs\rangle}{96\pi^2}\int_0^1 dy \left( \frac{1}{y}+\frac{1}{1-y} \right)\delta\left(s-\widetilde{m}_c^2\right) -t\frac{\langle\bar{s}s\rangle\langle\bar{s}g_s\sigma Gs\rangle}{192\pi^2}\int_{y_i}^{y_f} dy \, ,
\end{eqnarray}

\begin{eqnarray}
\rho_{10}(s)&=&-\frac{m_c^2\langle\bar{s}g_s\sigma Gs\rangle^2}{192\pi^2T^6}\int_0^1 dy \, \widetilde{m}_c^4 \, \delta \left( s-\widetilde{m}_c^2\right)
\nonumber\\
&&-t\frac{\langle\bar{s}g_s\sigma Gs\rangle^2}{64\pi^2}\int_0^1 dy \, y(1-y)\, \left(1+\frac{2 \widetilde{m}_c^2}{3T^2}+\frac{\widetilde{m}_c^4}{6T^4} \right)\delta \left( s-\widetilde{m}_c^2\right)\nonumber\\
&&+\frac{m_c^4\langle\bar{s}s\rangle^2}{216T^4}\langle\frac{\alpha_sGG}{\pi}\rangle\int_0^1 dy  \left\{ \frac{1}{y^3}+\frac{1}{(1-y)^3}\right\} \delta\left( s-\widetilde{m}_c^2\right)\nonumber\\
&&+t\frac{m_c^2\langle\bar{s}s\rangle^2}{432T^2}\langle\frac{\alpha_sGG}{\pi}\rangle\int_0^1 dy  \left\{ \frac{1-y}{y^2}+\frac{y}{(1-y)^2}\right\} \delta\left( s-\widetilde{m}_c^2\right)\nonumber\\
&&-\frac{m_c^2\langle\bar{s}s\rangle^2}{72T^2}\langle\frac{\alpha_sGG}{\pi}\rangle\int_0^1 dy  \left\{ \frac{1}{y^2}+\frac{1}{(1-y)^2}\right\} \delta\left( s-\widetilde{m}_c^2\right)\nonumber\\
&&+\frac{m_c^2\langle\bar{s}g_s\sigma Gs\rangle^2}{384 \pi^2T^4} \int_0^1 dy  \left( \frac{1}{y}+\frac{1}{1-y}\right)  \widetilde{m}_c^2 \, \delta\left( s-\widetilde{m}_c^2\right)\nonumber\\
&&+t\frac{ \langle\bar{s}g_s\sigma Gs\rangle^2}{768\pi^2 }\int_0^1 dy \left( 1+\frac{\widetilde{m}_c^2}{T^2}\right)\delta \left( s-\widetilde{m}_c^2\right)\nonumber\\
&&-\frac{m_c^2\langle\bar{s} s\rangle^2}{216 T^6}\langle\frac{\alpha_sGG}{\pi}\rangle\int_0^1 dy \, \widetilde{m}_c^4 \, \delta \left( s-\widetilde{m}_c^2\right)\nonumber\\
&&-t\frac{ \langle\bar{s} s\rangle^2}{72}\langle\frac{\alpha_sGG}{\pi}\rangle\int_0^1 dy \,y(1-y)\left(1+\frac{2\widetilde{m}_c^2}{3T^2} +\frac{\widetilde{m}_c^4}{6T^4} \right) \delta \left( s-\widetilde{m}_c^2\right)\, ,
\end{eqnarray}
 $\int dydz=\int_{y_i}^{y_f}dy \int_{z_i}^{1-y}dz$, $y_{f}=\frac{1+\sqrt{1-4m_c^2/s}}{2}$,
$y_{i}=\frac{1-\sqrt{1-4m_c^2/s}}{2}$, $z_{i}=\frac{y
m_c^2}{y s -m_c^2}$, $\overline{m}_c^2=\frac{(y+z)m_c^2}{yz}$,
$ \widetilde{m}_c^2=\frac{m_c^2}{y(1-y)}$, $\int_{y_i}^{y_f}dy \to \int_{0}^{1}dy$, $\int_{z_i}^{1-y}dz \to \int_{0}^{1-y}dz$ when the $\delta$ functions $\delta\left(s-\overline{m}_c^2\right)$ and $\delta\left(s-\widetilde{m}_c^2\right)$ appear.

\section*{Acknowledgements}
This  work is supported by National Natural Science Foundation,
Grant Numbers 11375063, the Fundamental Research Funds for the
Central Universities,  and Natural Science Foundation of Hebei province, Grant Number A2014502017.
The author would like to thank Prof. T. Huang for suggesting this subject.

\end{document}